\documentclass[11pt, twocolumn]{article}
\usepackage[a4paper,
            left=0.5in,
            right=0.5in,
            top=0.6in,
            bottom=0.8in,
            footskip=0.4in]{geometry}
\usepackage[utf8]{inputenc}

\usepackage{amsmath,amssymb,amsfonts,amsthm,color}

\usepackage{graphicx}
\usepackage{subfigure}

\usepackage[usenames]{xcolor}

\usepackage{caption}
\usepackage[normalem]{ulem} 

\usepackage{url}
\usepackage{hyperref}
\hypersetup{
  colorlinks   = true, 
  urlcolor     = blue, 
  linkcolor    = black, 
  citecolor   = black 
}

\usepackage[capitalise]{cleveref}

\usepackage{dsfont}
\usepackage{lipsum}

\usepackage[numbers,sort&compress]{natbib}
\let\OLDthebibliography\thebibliography
\renewcommand\thebibliography[1]{
  \OLDthebibliography{#1}
  \setlength{\parskip}{2pt}
  \setlength{\itemsep}{0pt plus 0.3ex}
}

\allowdisplaybreaks

\newcommand{\abs}[1]{\ensuremath{\left\lvert {#1} \right\rvert}}

\newcommand\refta[1]{Tab.~\ref{#1}}

\newcommand\citere[1]{Ref.~\cite{#1}}
\newcommand\citeres[1]{Refs.~\cite{#1}}

\def\reffi#1{\mbox{Fig.~\ref{#1}}}

\newcommand{\TB}[1]{{\color{black}#1}}

\newcommand{\GW}[1]{{\color{black}#1}}

\newcommand{\ps}[1]{{\color{black}#1}}

\newcommand{\fa}[1]{{\color{black}#1}}

\newcommand{\psalt}[1]{{\color{black}#1}}
\newcommand{\faalt}[1]{{\color{black}#1}}

\newcommand{\madgraph}{{\sc{MadGraph5\textunderscore aMC@NLO}}}
\newcommand{\higgsbounds}{{\sc{HiggsBounds}}}
\newcommand{\higgstools}{{\sc{HiggsTools}}}

\newcommand{\chel}{\ensuremath{{c_{hel}}}}
\newcommand{\chan}{\ensuremath{{c_{han}}}}

\usepackage{abstract}

\author{thomas.biekoetter@kit.edu}

\date{\today}

\usepackage{titlesec}
\titleformat{\section}
{\normalfont\bfseries}{\thesection}{1em}{}
\titleformat{\subsection}
{\normalfont\bfseries}{\thesubsection}{1em}{  }

\begin{document}

\def\thefootnote{\fnsymbol{footnote}}

\twocolumn[
\begin{@twocolumnfalse}
\begin{flushright}
\footnotesize
  IFT–UAM/CSIC-24-187 ~~ 
  DESY-25-023
\end{flushright}

\begin{center}
{\large
\textbf{\TB{
Top-quark spin correlations as a tool to distinguish
pseudoscalar $A \to ZH$\\[0.4em] and scalar $H \to ZA$ signatures
in $Z t \bar t$ final states at the LHC
}}
}
\vspace{0.4cm}

Francisco Arco$^{1*}$,
Thomas Biekötter$^{2\dagger}$,
Panagiotis Stylianou$^{1\ddagger}$
and
Georg Weiglein$^{1,3\mathsection}$\\[0.8em]

{\small

 $^1${\textit{
   Deutsches Elektronen-Synchrotron DESY,
     Notkestr.~85, 22607 Hamburg, Germany
  }}\\[0.4em]

  $^2${\textit{
   Instituto de F\'isica Te\'orica UAM/CSIC,
   Calle Nicolás Cabrera 13-15,
   Cantoblanco, 28049, Madrid, Spain
 }}\\[0.4em]

 $^3${\textit{
   II. Institut für Theoretische Physik, Universität Hamburg,
    Luruper Chaussee 149, 22761 Hamburg, Germany\\[0.8em]
  }}

}

\begin{abstract}
Both ATLAS and CMS have recently
performed the first searches for a heavy new
spin-0 resonance decaying
into a lighter new spin-0 resonance and a $Z$ boson, where
the lighter spin-0 resonance subsequently decays
into $t \bar t$ pairs.
These searches are of particular
interest to probe Two Higgs doublet model~(2HDM)
parameter space regions that predict a strong
first-order electroweak phase transition.
In the absence of CP violation, the investigated decay is
possible if the lighter
and the heavier spin-0 particles have opposite
CP parities. The analysis techniques employed
by ATLAS and CMS do not distinguish
between the two possible signatures $A \to ZH$
and $H \to ZA$,
where $A$ and $H$ denote CP-odd and CP-even Higgs bosons,
respectively, if 
both signals are predicted to have the same total
cross sections.
We demonstrate
the capability of angular variables
\GW{that are}
sensitive to spin correlations of the top quarks
to differentiate between $A \to ZH$ and $H \to ZA$ decays,
even in scenarios where both signals
possess identical total
cross sections. 
Focusing on masses of 600~GeV and 800~GeV
as a representative 2HDM benchmark, 
we find that a distinction between
the two possible channels is possible with high
significance 
with the anticipated data from the high-luminosity LHC,
if the invariant mass distribution of the
$t \bar t$ system is further binned in
angular variables defined by the direction of flight
of the leptons produced in the top-quark decays.
Moreover, we find a 
\GW{moderate}
gain in experimental sensitivity due to
the improved background rejection for both signals. 
\end{abstract}

\end{center}
\end{@twocolumnfalse}
]

\section{Introduction}
\label{sec:introduction}

\textcolor{white}{
 {\fnsymbol{footnote}}\stepcounter{footnote}
 \footnotetext{francisco.arco@desy.de}
 {\fnsymbol{footnote}}\stepcounter{footnote}
 \footnotetext{thomas.biekoetter@desy.de}
 {\fnsymbol{footnote}}\stepcounter{footnote}
 \footnotetext{panagiotis.stylianou@desy.de}
 {\fnsymbol{footnote}}\stepcounter{footnote}
 \footnotetext{georg.weiglein@desy.de}
  }
\vspace*{-1.2em}
  
\renewcommand{\thefootnote}{\arabic{footnote}}
\setcounter{footnote}{0} 

\noindent 
In 2012 the LHC discovered a Higgs boson which,
at the current level of experimental precision,
behaves in agreement with the predictions
of the Standard Model~(SM)~\cite{CMS:2022dwd,
ATLAS:2022vkf}.
While the SM predicts only one Higgs boson,
theories beyond the SM~(BSM) often contain
more than one fundamental spin-0 particle.
Consequently, the search for additional
Higgs bosons is one of the prime \GW{tasks}
of the current and future LHC
programme.

Most searches for additional Higgs bosons
focus on the production of one BSM resonance.
However, BSM theories that contain additional
Higgs fields \GW{that are} charged under the electroweak~(EW)
gauge \TB{symmetry} predict more than one BSM Higgs boson.
Such BSM theories
have the potential to
resolve some of the most pressing open questions
that remain unanswered in the SM,
e.g.\ \GW{extended Higgs sectors can provide an explanation of 
the observed} matter-antimatter asymmetry 
via EW baryogenesis~\cite{Kuzmin:1985mm},
and additional neutral
scalar particles can be stable and account
for the observed cosmological abundance
of dark matter.
Consequently, during Run~2 at 13~TeV, ATLAS
and CMS also performed
searches for signals in which two BSM spin-0
resonances are involved. 
These searches mostly comprise signatures with
a heavy BSM resonance decaying into a lighter
BSM resonance and either a 125~GeV
Higgs boson
or a massive gauge
boson~\cite{CMS:2024phk}.

Among \TB{these,} 
searches for a neutral 
spin-0 particle decaying into another
neutral 
spin-0 particle and a $Z$-boson have gathered
\GW{significant}
attention~\cite{Biekotter:2021ysx,
Goncalves:2022wbp,Biekotter:2022kgf,
MammenAbraham:2022yxp,Biekotter:2023eil}.
This 
\GW{search channel}
has been
identified as a \GW{``smoking-gun''} signature
for a first-order EW phase transition in the
Two Higgs doublet
model~(2HDM)~\cite{Dorsch:2014qja,
Dorsch:2016tab,
Biekotter:2022kgf,Biekotter:2023eil}.
A sufficiently strong EW phase transition is
a required ingredient for the realisation
of EW baryogenesis~\cite{Kuzmin:1985mm,
Cline:1996mga,
Fromme:2006cm,Basler:2016obg},
and it leads to the production of
a primordial gravitational-wave background
that might be in reach of future
space-based gravitational-wave
detectors~\cite{Dorsch:2016nrg,
Goncalves:2021egx,
Biekotter:2022kgf}.

\TB{Assuming CP conservation,
t}he 2HDM predicts a second CP-even
Higgs boson~$H$ and a CP-odd
Higgs boson~$A$.
A strong EW phase transition typically requires
a sizeable mass splitting between these
two states~\cite{Dorsch:2016tab}, and
(depending on their
mass hierarchy) either
the decay $H \to ZA$ or the decay
$A \to ZH$ can be kinematically allowed.
\TB{Furthermore,
since the top quark has the largest Yukawa coupling,
its interactions are 
\GW{well-suited for providing}
the CP-violating
source term that generates the baryon asymmetry.
As a consequence,
small values
of $\tan\beta$ are preferred for EW
baryogenesis~\cite{Fromme:2006wx}, and}
the dominant decay modes for the lighter
resonance in the 2HDM are
$A/H \to t \bar t$
if its mass exceeds the di-top threshold,
giving rise to $Z t \bar t$ final states.
Notably,  parameter points facilitating the decay $A \to ZH$
are more
favourable for a successfully realisation
of EW baryogenesis compared to the ones
featuring the $H \to ZA$ decay~\cite{Dorsch:2014qja,
Basler:2017uxn,Basler:2016obg,Dorsch:2017nza,
Su:2020pjw}.
It \GW{will} therefore \GW{be} \TB{crucial}
to be able to experimentally distinguish
between the two decay modes if 
a 
signal 
\GW{in the ``smoking gun'' searches will be}
observed at
the LHC in the future.

Searches for this signal in $Z t \bar t$
final states have been recently
performed for the first time by both ATLAS
and CMS, using the Run~2 data collected at
13~TeV~\cite{ATLAS:2023zkt,CMS-PAS-B2G-23-006}.
The resulting experimental limits have
been exploited in \citere{Biekotter:2023eil}
\GW{by demonstrating that they}
exclude substantial parts of the 2HDM
parameter space 
\GW{giving rise to}
a strong
first-order EW phase transition.
In both the ATLAS and the CMS analyses,
the applied experimental analyses
lack sensitivity to the
CP-properties of the BSM particles.
Therefore, a distinction
between the $A \to ZH$ and the $H \to ZA$
signatures is not possible if both signals
predict the same total cross section.
In this work, we propose to use
angular variables in $Z t \bar t$ final states
in order to distinguish between
$A \to ZH$ and $H \to ZA$ signals where the lighter
BSM particle decays into a pair of top quarks\TB{,
which subsequently decay leptonically.}

Our study demonstrates that sensitivity to
the CP-properties of the BSM particles
can be achieved by exploiting the dependence
of the $t \bar t$ invariant mass distribution
($m_{t \bar t}$) on angular variables defined
in terms of the direction of flight of the leptons
produced in the leptonic decay of the
top quarks.\footnote{An analysis of
the CP properties of a BSM resonance decaying
into a $Z$ boson and the 125~GeV Higgs boson,
the latter assumed to be purely CP even,
can be found in \citere{Bauer:2016zfj}.}
By considering the angular
correlations of the leptons, we show
that distinguishing between 
\GW{the two}
signatures
is feasible 
with the anticipated
3000/fb of data collected during
the high-luminosity phase of the LHC,
even if they possess the same total
cross section. It is worth noting that the
relevant angular variables have
been previously employed in searches for
\GW{a single new particle}
decaying into top-quark pairs
at both the Tevatron~\cite{D0:2011kcb}
and the LHC~\cite{CMS:2019pzc,ATLAS:2024vxm,
CMS-PAS-HIG-22-013}, \GW{see also \citere{Anuar:2024qsz}.}
\GW{We demonstrate here the potential of extending such an}
experimental strategy 
\GW{based on angular variables to}
signatures involving two BSM particles.

In addition to 
\GW{obtaining a}
discrimination power
between the two potential signals, we
demonstrate that a refined
analysis technique utilizing angular variables
\TB{also improves} the
signal-background discrimination, \TB{and
thus the overall experimental sensitivity}.
Here it should be noted that
our approach relies on the
leptonic \GW{decays of both} top quarks, in contrast
to the experimental analyses conducted
by ATLAS~\cite{ATLAS:2023zkt}
and CMS~\cite{CMS-PAS-B2G-23-006},
which \fa{only} incorporate the semi-leptonic and
the fully hadronic decays, respectively.
As a result, while there exists potential for
increased background rejection, this potential
\GW{of our proposed analysis comes at the cost of a reduced number}
of signal
events resulting from the requirement of
leptonically decaying top quarks.
\TB{On the other hand,} 
\GW{the angular information obtained from
the leptons} in the signal regions where
both top quarks decay leptonically 
\GW{can be}
combined 
with the experimental
information
gained from the 
signal regions
targeting semileptonically and
hadronically decaying $t \bar t$
pairs for which our approach is not 
\GW{directly}
applicable.\footnote{The
measurement of $t \bar t$ spin
correlations 
\GW{may also be}
feasible in the semileptonic channel with the data
collected during the high-luminosity phase
of the~LHC~\cite{Tweedie:2014yda,Dong:2023xiw}, 
\GW{relying}
on the possibility of charm-tagging~\cite{Stelzer:1995gc}.
However, we focus 
\GW{here}
on the fully leptonic channel
\TB{since it is the final state with the highest
sensitivity to the $t \bar t$ spin information,
see the discussion in \cref{sec:angular}.}}

The outline of our paper is as follows.
In \cref{sec:theoretical}
we introduce the theoretical
framework, including a brief description
of the 2HDM that we use as a theoretical
framework in order to benchmark our
results, and a detailed discussion of
the Monte Carlo simulation of the considered
process. In particular,
we discuss the implementation
of the angular variables in the
experimental analysis that are sensitive
to the spin correlations of the top-quark pairs.
In \cref{sec:numerical}
we present the results of
our simulation, focusing on the one
hand on the
discrimination between the signals and
the SM background,
and on the other hand on the discrimination
between the two potential signals.
We summarise 
\GW{our results}
and conclude 
in \cref{sec:conclusions}.

\section{Theoretical framework and simulation}
\label{sec:theoretical}

\begin{figure}[t]
\centering
\includegraphics[width=0.6\columnwidth]{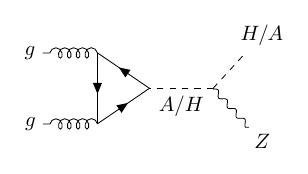}
\caption{Feynman diagram of the considered
signals with $H$ and $A$ being a CP-even
and a CP-odd spin-0 resonance, respectively.}
\label{fig:signaldiag}
\end{figure}

In order to demonstrate the 
\GW{application of the}
angular variables \GW{to} the $m_{t \bar t}$
distribution in the processes
$A \to ZH \to Z t \bar t$ and
$H \to ZA \to Z t \bar t$, we perform
a Monte-Carlo~(MC) simulation of both
signal\fa{s} and SM background.
\TB{The Feynman diagram for the 
\GW{production of the}
two signals
is depicted in \reffi{fig:signaldiag}.}
For our simulations
we choose the 2HDM as theoretical framework,
noting that our conclusions can be applied
more generally to any BSM theory with an extended
scalar sector featuring at least two
neutral extra Higgs bosons with opposite
CP charges (or potentially 
\GW{two CP-mixed states for the case}
of CP violation)
such that a coupling of the form $AHZ$ is present.

\subsection{The 2HDM}
\label{sec:2hdm}

The 2HDM~\cite{Lee:1973iz} augments the
scalar sector of the SM by
a second scalar SU(2) doublet field.
\GW{For the case of}
CP conservation, the 2HDM predicts
two CP-even Higgs bosons $h$ and $H$,
one CP-odd Higgs boson~$A$,
and a pair of charged Higgs bosons~$H^\pm$.
\fa{The angles $\alpha$ and $\beta$ \GW{diagonalise} the
CP-even and CP-odd \TB{scalar} sectors of the model, respectively.
In addition, $t_\beta  \equiv \tan\beta$ 
\GW{is given by} the ratio of the two Higgs doublet vacuum expectation values.}
Throughout this work,
the lighter CP-even Higgs boson~$h$ 
\GW{corresponds to}
the detected
Higgs boson at 125~GeV.
Parameter space regions that 
\GW{give rise to}
EW baryogenesis \GW{favour}
small values of $t_\beta$ \GW{(see the discussion above) and}
the alignment
limit \fa{(defined by $\cos(\beta-\alpha)=0$),} in which $h$ resembles \GW{the} Higgs boson
predicted by the SM, \GW{as well as}
a sizeable mass splitting between the
pseudoscalar $A$ and the
second CP-even scalar $H$.

\begin{table}[t]
\centering
\renewcommand{\arraystretch}{1.2}
\begin{tabular}{l||r|r}
 & $\textrm{BP}_{H \to ZA}$ & $\textrm{BP}_{A \to ZH}$ \\
\hline
\hline
$\tan\beta$ & 1.14 & 1.50 \\
$\cos(\beta - \alpha)$ & 0 & 0 \\
$m_h/$GeV & 125 & 125 \\
$m_H/$GeV & 800 & 600 \\
$m_A/$GeV & 600 & 800 \\
$m_{H^\pm}/$GeV & 800 & 800 \\
$M/$GeV & 600 & 600 \\
\hline
$\mathrm{BR}(H \to t \bar t)$ & 71\% & 99\% \\
$\mathrm{BR}(A \to t \bar t)$ & 99\% & 63\%\\
$\mathrm{BR}(H \to ZA)$ & 29\% & -- \\
$\mathrm{BR}(A \to ZH)$ & -- & 37\% \\
$\Gamma_H/m_H$ & 4.3\% & 1.5\% \\
$\Gamma_A/m_A$ & 3.5\% & 3.3\% \\
$\sigma(gg \to H)/$pb & 0.35 & 0.89 \\
$\sigma(gg \to A)/$pb & 2.43 & 0.27 \\
\end{tabular}
\renewcommand{\arraystretch}{1.0}
\caption{Definitions of the benchmark points
$\textrm{BP}_{H \to ZA}$ and $\textrm{BP}_{A \to ZH}$
and predicted branching ratios,
total widths and gluon-fusion production
cross sections at the LHC \TB{with $\sqrt{s} = 13$~TeV}.
}
\label{tab:benchis}
\end{table}

Our analysis targets signal cross sections
that are 
\GW{compatible with the limits from}
the LHC
searches performed during Run~2, but
which lie in reach of the high-luminosity
LHC~(HL-LHC) with an anticipated
integrated luminosity of
$3000/\mathrm{fb}$. For ease of comparison we 
use a centre-of-mass energy of $\sqrt{s}=13$ TeV
also for the HL-LHC.
As a representative 2HDM benchmark point 
we choose masses of 600~GeV for the lighter
and 800~GeV for the heavier BSM resonance,
respectively. The mass of the charged
Higgs bosons is set to be equal to the
mass of the heavier BSM state,
$m_{H^\pm} = 800$~GeV, to satisfy constraints
from electroweak precision
observables~\cite{Gerard:2007kn}.
We also assume the alignment
limit\fa{, i.e.~}$\cos (\beta - \alpha) = 0$.
Regarding $t_\beta$, we use two different
values depending on the mass hierarchies
of $H$ and $A$ in order to predict
the same total cross section
for the two possible signals.
We use a value
of $t_\beta = 1.5$ for the case
$m_A = 800$~GeV and $m_H = 600$~GeV,
and $t_\beta = 1.14$ for the case
$m_H = 800$~GeV and $m_A = 600$~GeV,
such that for both signals
we find $\sigma[pp \to A(H) \to Z \,
H(A) \to Z t \bar t] = 0.1~\mathrm{pb}$.
\TB{Here the production cross sections of
$H$ and $A$ were computed at NNLO in QCD using
\higgstools~\cite{Bahl:2022igd}
(see also \GW{the} discussion in \cref{sec:signalsim}),
and their branching ratios were computed using
\textsc{HDECAY}~\cite{Djouadi:1997yw,Djouadi:2018xqq},
including state-of-the-art QCD corrections.}
The remaining 2HDM parameter $m_{12}^2$
is not relevant for the considered
signature. To fix $m_{12}^2$,
in our analysis we set
the BSM mass scale $M^2 = m_{12}^2 /
(\sin\beta \, \cos\beta)$ equal
to the mass of the lighter BSM Higgs
boson, $M = 600$~GeV, in order to comply
with theoretical constraints from
vacuum stability~\cite{Barroso:2013awa,Hollik:2018wrr}
and perturbative
unitarity~\cite{Cacchio:2016qyh}
which we checked using
\textsc{thdmTools}~\cite{Biekotter:2023eil}.
\TB{The values of the free parameters for
the two considered benchmark scenarios
are summarised in \refta{tab:benchis},
where we also show the relevant branching ratios,
the total widths
and the cross sections of the neutral BSM scalars.

Using the \higgsbounds~\cite{Bechtle:2008jh,
Bechtle:2011sb,Bechtle:2013wla,Bechtle:2020pkv}
module contained in \higgstools~\cite{Bahl:2022igd},
we verified that the
two benchmark points pass \GW{the} LHC cross section
limits from
searches for $H^\pm \to tb$~\cite{CMS:2020imj,ATLAS:2021upq}
and from searches for $A/H \to t \bar t$ in
$t \bar t t \bar t$ final
states~\cite{CMS:2019rvj,ATLAS:2024jja}.
The benchmark points are also compatible with
the limits on $H t \bar t$ and
$A t \bar t$ couplings
from searches for $A/H \to t \bar t$ in the di-top
final state obtained by CMS utilising the
first-year Run~2 data~\cite{CMS:2019pzc}.
Recently, both ATLAS~\cite{ATLAS:2024vxm}
and CMS~\cite{CMS-PAS-HIG-22-013}
reported preliminary results of searches in the di-top
final state including the full Run~2 dataset.
Due to the large interference effects between the
signal and the QCD background the resulting limits
depend on the width of the new particle.
No limits are presented for a relative width
of about 3.5\% that we find for the lighter resonance
in our benchmark points, see \cref{tab:benchis}.
Assuming that the
limits given for a relative width of 5\% are
approximately applicable
to our benchmark points, the \GW{limits from the}
new searches would
be in tension with our benchmark points at
about \GW{the $2 \, \sigma$} 
level.
However, since we are mainly interested in the
improvement of the searches in the $Z t \bar t$
final state 
\GW{that can be achieved by exploiting}
top-quark spin correlations,
and not in a comparison between searches in
different final states, we stick to our benchmark
points.
We also stress that the cross sections for
the $gg \to A \to ZH$ and $gg \to H \to ZA$ signals
decrease more slowly with \GW{increasing} $t_\beta$ compared
to the direct production of the lighter state
$gg \to H$ and $gg \to A$, respectively,
see \GW{the} discussion below.
Hence, for $t_\beta$ values larger than the ones
considered in our benchmark points, the lighter state
could be detected first via its production from
the decay of a heavier BSM resonance as
in the channel investigated
here 
\GW{rather than}
via its direct production and
searches using the $t \bar t$ final
state~\cite{Biekotter:2023eil}.}

\begin{figure}[t]
\centering
\includegraphics[width=0.98\columnwidth]{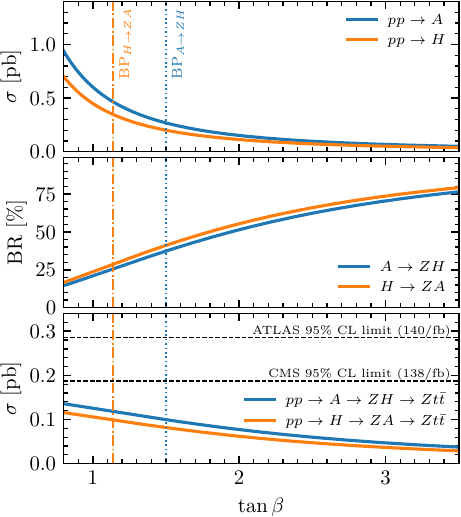}
\caption{Top: gluon-fusion production cross section
at 13~TeV for the heavier BSM spin-0 resonance
\GW{(with a mass of 800~GeV)}
as a function of $t_\beta$.
\GW{Centre}: branching ratios for the
heavy spin-0 resonance decaying into the lighter
spin-0 resonance 
\GW{(with a mass of 600~GeV)}
and a $Z$-boson as a function
of $t_\beta$.
Bottom: Signal cross sections contributing to
$Z t \bar t$ production as a function of $t_\beta$.
\TB{The horizontal black dashed lines show the current
experimental 95\% CL limits 
\GW{from}
ATLAS~\cite{ATLAS:2023zkt}
and CMS~\cite{CMS-PAS-B2G-23-006}
using an integrated luminosity of
140/fb and 138/fb, respectively.}
In all plots the \GW{orange and blue vertical}
\TB{lines correspond to the
benchmark points $\mathrm{BP}_{H \to ZA}$ and
$\mathrm{BP}_{A \to ZH}$, respectively, which
are defined in \refta{tab:benchis}.}
}
\label{fig:entryplot}
\end{figure}

\TB{W}e show in
\cref{fig:entryplot} the total cross sections
\GW{for}
the heavier BSM resonance (top), the branching ratios
for the $A \to ZH$ and $H \to ZA$ decays (middle), and the
total signal cross section contributing to
$Z t \bar t$ production (bottom) as a function
of $t_\beta$\TB{, with all other parameters fixed
as shown in \refta{tab:benchis}}.
The gluon fusion production is dominated
by the contribution from the top-quark loop. Since the
absolute values of the
couplings of $H$ and $A$ to the top quark scale with
a factor of $1 / t_\beta$, the cross sections shown
in the \fa{top} plot
are approximately proportional to
$1 / t_\beta^2$. The 
\GW{decrease of}
the production cross
section of the heavier BSM resonance with increasing
values of $t_\beta$ is partially compensated by
an increase of $\mathrm{BR}( A \to ZH )$ and
$\mathrm{BR} ( H \to ZA )$ with increasing value
of $t_\beta$, as shown in the plot in the middle.
In both benchmark scenarios, the lighter BSM resonance dominantly
decays via $H/A \to t \bar t$ with a branching ratio
of more than 99\% for the small values of $t_\beta$
relevant here.
As a consequence, the final signal cross sections
show an approximately linear dependence on $1 / t_\beta$,
as is visible in the bottom plot\TB{, where we
also indicate with the \GW{horizontal} black dashed
lines the current 95\% CL cross-section limits
found by ATLAS~\cite{ATLAS:2023zkt}
and CMS~\cite{CMS-PAS-B2G-23-006} including 
\GW{140/fb and 138/fb, respectively,} collected
during Run~2.}

\subsection{Angular variables}
\label{sec:angular}
To gain information on the CP nature of
the BSM resonances, we propose to utilize
the spin correlations of the final state
$t \bar t$ pair.
The production density matrix of two top quarks can be written in terms of the Pauli matrices $\sigma$ as \cite{Bernreuther:2015yna}
\begin{equation}
    R \propto A \, \mathds{1}\otimes\mathds{1} + B_i^+ \sigma^i \otimes \mathds{1} + B_i^- \mathds{1}\otimes\sigma^i + C_{ij} \sigma^i \otimes \sigma^j \;,
\end{equation}
where $A$ and the vector $\vec{B}^{\pm}$ 
\GW{arise from}
the parton level kinematics and the polarisations
of the di-top state, respectively.
The spin correlations of the top and anti-top
quarks are encoded in the 
\GW{matrix $C$.
They}
are commonly extracted
experimentally by evaluating observables in
an orthonormal basis $(\hat{k}, \hat{r}, \hat{n})$.
The coordinate $\hat{k}$ is defined as the unit
vector of the top-quark direction in the
zero-momentum frame~(ZMF), \GW{which is}
\TB{equal to the \GW{centre}-of-mass frame of
the $t \bar t$ system.}
Taking $\hat{p}$ as the direction of flight
of one of the incoming protons,
the scattering angle of the top quark is
given by $\cos\theta_t = \hat{p} \cdot \hat{k}$,
which can be used to obtain the unit vector
\begin{equation}
    \hat{n} = \frac{\textrm{sign}{(\cos\theta_t})}{\sin\theta_t} (\hat{p} \times \hat{k})\;.
\end{equation}
We define the remaining coordinate as
$\hat{r} = - \hat{n} \times \hat{k}$.
Assuming fully leptonic decays of 
both top quarks, the leptons are boosted
first to the di-top ZMF frame and subsequently
to their respective parent top-quark ZMF.
Their directions of flight are denoted as
\GW{$\hat\ell^+$ and $\hat\ell^-$}.
The normalised angular distributions can then be
written in terms of $\GW{\vec{B}}^\pm$ and $C$ as 
\begin{equation}
    \begin{split}
    \frac{1}{\sigma}\frac{d \sigma}{d \Omega^+ d \Omega^-} = \frac{1}{(4 \pi)^2} ( 1 &+ \kappa_\ell \vec{B}^+ \cdot \hat{\ell}^+ + \kappa_\ell \vec{B}^- \cdot \hat{\ell}^- 
     \\  &- \kappa_\ell^2 \hat{\ell}^+ \cdot C \cdot \hat{\ell}^- ) 
    \end{split}
\end{equation}
for solid angles $d \Omega^\pm$. 
We assume a spin analysing power of $\kappa_\ell = 1$
for leptons, which is the case
at tree-level (higher orders effects are of 
\GW{relative} order 
$10^{-3}$~\cite{Czarnecki:1990pe} \GW{or} smaller).
Quarks have spin analysing powers smaller than unity
and receive larger corrections from higher QCD orders,
rendering the fully-leptonic channel the easiest case
for extracting the top-quark
spin-correlations~\cite{Brandenburg:2002xr}.

Choosing a reference axis $\hat{a} \in \{\hat{k},\hat{r},\hat{n}\}$, the angle of the lepton and the axis is given by $\cos\theta_{\hat{a}}^\pm = \pm \hat{\ell}^\pm \cdot \hat{a}$.
Allowing for different axes $\hat{a}, \hat{b} \in \{\hat{k},\hat{r},\hat{n}\}$, the differential cross section for a choice of axes after integrating over azimuthal angles is then given by
\begin{align}
    \frac{1}{\sigma}\frac{d \sigma}{d \cos\theta^+_{\hat{a}} d \cos\theta^+_{\hat{b}}} = \frac{1}{4} ( 1 &+ B^+_{\hat{a}} \cos\theta^+_{\hat{a}} + B^-_{\hat{a}} \cos\theta^-_{\hat{a}} \notag
    \\  &- C_{\hat{a}\hat{b}} \cos\theta^+_{\hat{a}} \cos\theta^-_{\hat{b}}) \; ,
\end{align}
where we have written the vectors $\vec{B}^\pm$
and the matrix $C$ in terms of their components.
The connection of $\cos\theta^+_{\hat{a}}
\cos\theta^-_{\hat{b}}$ to the spin correlations of
the $t\bar{t}$ system renders them 
\GW{well-suited}
observables
to distinguish between 
\GW{cases} of different parities.
\GW{This can be employed}
for example (as considered here)
\GW{to distinguish}
whether the top-quark pair originated from
a scalar or pseudoscalar state.
\GW{Similarly} to~\GW{\citeres{Bernreuther:2004jv,
Aguilar-Saavedra:2022uye,
Rubenach:2023opp,CMS-PAS-HIG-22-013}} we use the observables
\begin{align}
    \chel &= - \cos\theta^+_{\hat{k}} \cos\theta^-_{\hat{k}} - \cos\theta^+_{\hat{r}} \cos\theta^-_{\hat{r}} - \cos\theta^+_{\hat{n}} \cos\theta^-_{\hat{n}} \;, \nonumber \\
    \chan &=  \cos\theta^+_{\hat{k}} \cos\theta^-_{\hat{k}} - \cos\theta^+_{\hat{r}} \cos\theta^-_{\hat{r}} - \cos\theta^+_{\hat{n}} \cos\theta^-_{\hat{n}} \;,
    \label{eq:chanhel}
\end{align}
which are sensitive to the CP-nature of \GW{the} state producing
the top-quark pair.\footnote{The parameter
$\chel$ is called $D$ in \citere{Bernreuther:2004jv},
and $\chan$ is called $D_3$ in
\citere{Aguilar-Saavedra:2022uye}.}
The diagonal spin correlation coefficients that
contribute to the $\chel$ and $\chan$ observables have
been studied for the $t \bar{t} Z$ channel in \citere{Ravina:2021kpr}.
\GW{They} obtain different values compared to the $t \bar{t}$ channel (without an emitted $Z$ boson)
for the SM. Most importantly, in $t \bar{t} Z$ the diagonal spin correlation 
coefficients have the
opposite sign compared to $t \bar{t}$ which implies that the
same should be true for $\chel$.
We investigate this further including the effects from additional (pseudo)scalars in 
\cref{sec:background,sec:signalsim}.

\subsection{Monte Carlo simulation}
\label{sec:simulation}
To study the impact on the $\chel$ and $\chan$
observables from additional scalar states in
the $t \bar{t} Z$ channel, we perform a
numerical MC
simulation 
using~\madgraph~\cite{Alwall:2014hca,Frederix:2018nkq}.
We use {\textsc{FeynRules}}~\cite{Christensen:2008py,Alloul:2013bka}
to extend the SM model with the additional interactions
of interest 
that couple
the scalar and pseudoscalar to
the top quark and the $Z$ boson,
\begin{equation}
	\label{eq:efflang}
	{\cal{L}} \supset - \frac{m_t}{v t_\beta} \bar{t} (H + i A \gamma_5) t - \frac{e}{2 s_W c_W} (H \partial_\mu A - A \partial_\mu H) Z^\mu \;,
\end{equation}
where $s_W$, $c_W$ denote the sine and cosine of the
weak mixing
angle, \GW{$v \approx 246$}~GeV is the SM Higgs vacuum expectation value, and $e$ is
the electric charge. The effective interactions between
the (pseudo)scalar and the gluon field are
also introduced,
\begin{equation}
	\label{eq:effggF}
	{\cal{L}} \supset \frac{\alpha_S}{8 \pi v} \left[ {\cal{F}}_H(\tau) H G^a_{\mu\nu} G^{a\mu\nu} + i {\cal{F}}_A (\tau) A G^{a}_{\mu\nu} \widetilde{G}^{a \mu\nu}\right]\;,
\end{equation}
\GW{where}
the strong coupling constant \GW{is} denoted by $\alpha_S$.
The interactions are exported as a
{\textsc{UFO}}~\cite{Degrande:2011ua,Darme:2023jdn}
model file with an additional pseudoscalar~$A$ and
scalar~$H$.
\GW{The {\textsc{UFO}} file}
is extended to include the
momentum-dependent form factors arising from the
top-quark triangle loop~\cite{Spira:1995rr}
\begin{align}
	\label{eq:formfactors}
		{\cal{F}}_H (\tau) &= \frac{1}{\tau^2} \left( \tau + (\tau - 1) f(\tau) \right) \;,\\ 
		{\cal{F}}_A (\tau) &= - \frac{1}{\tau} f(\tau) \;,
\end{align}
where $\tau = \hat{s} / (4 m_t^2)$. The function $f(\tau)$ is given by
\begin{equation}
	\label{eq:loopf}
	f(\tau) = \begin{cases}
					\arcsin^2 \left(\sqrt{\tau}\right) & \tau \leq 1\;, \\
					-\frac{1}{4} \left[ \log{\frac{1+\sqrt{1-\tau^{-1}}}{1 - \sqrt{1-\tau^{-1}}}} - i \pi \right]^2 & \tau > 1\;.
				\end{cases}
\end{equation}
We have cross-checked our matching of the effective vertex to the triangle loop contribution using {\sc{FeynArts}}~\cite{Kublbeck:1990xc,Hahn:2000kx} and {\sc{FormCalc}}~\cite{Hahn:2001rv}. 

\subsubsection{Background}
\label{sec:background}

\begin{figure*}[t]
\centering
\includegraphics[width=0.62\columnwidth]{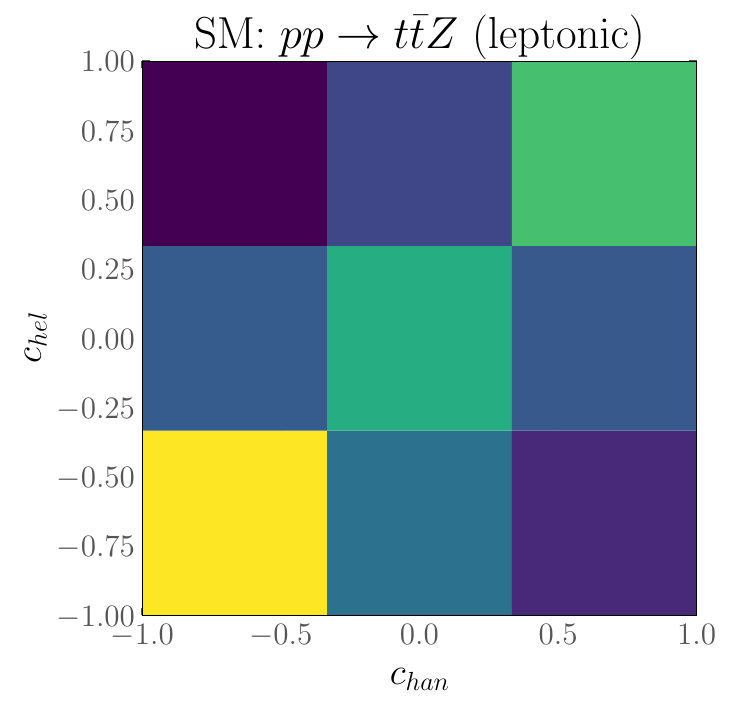}\hspace{-0.02\columnwidth}
\includegraphics[width=0.62\columnwidth]{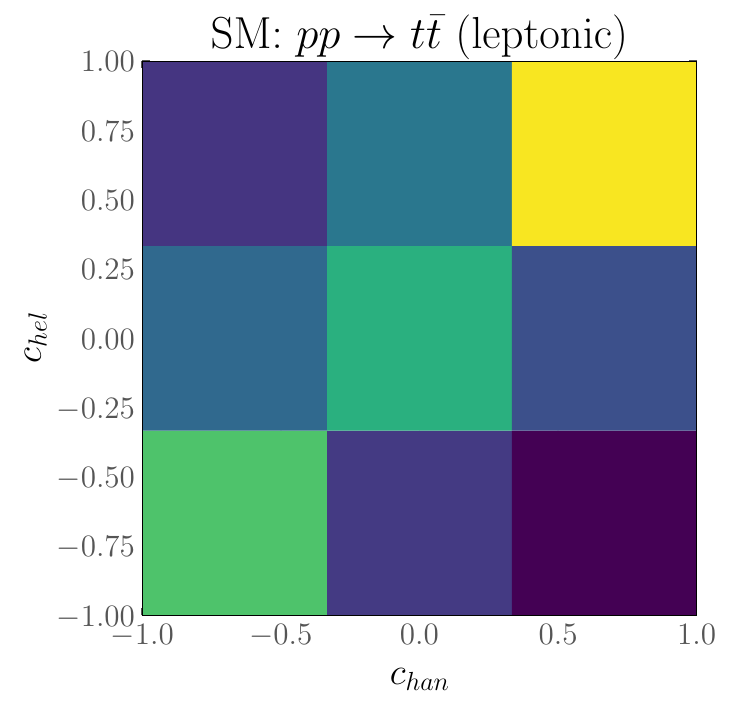}\hspace{-0.02\columnwidth}
\includegraphics[width=0.778\columnwidth]{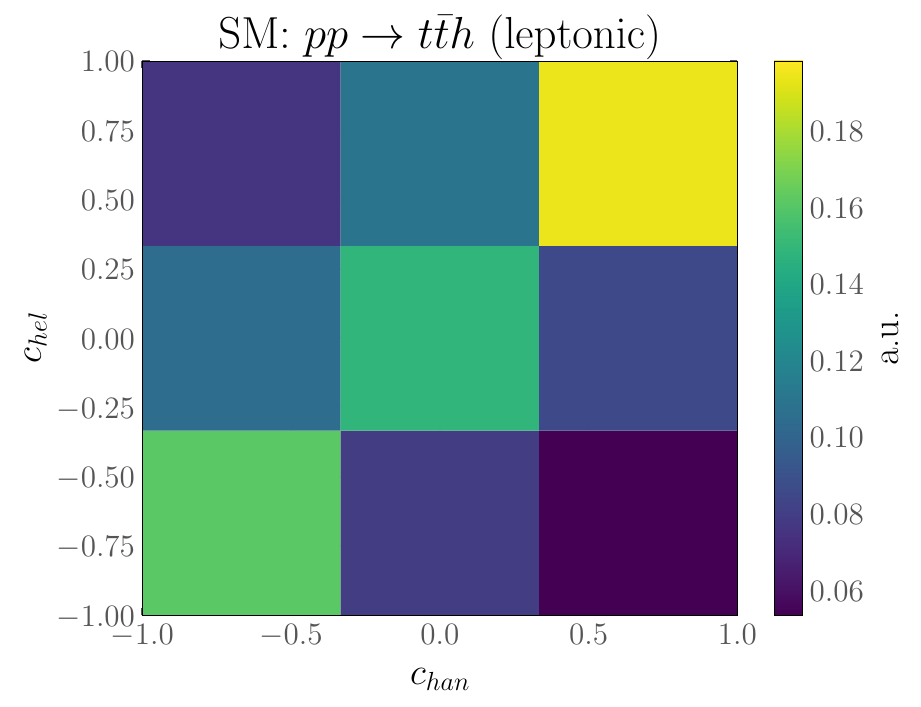}
\caption{Two-dimensional distributions in the $\chel$-$\chan$ plane for different SM channels. Our process of interest $t \bar{t} Z$ (left) has a $\chel$ value with opposite sign compared to $t \bar{t}$ (middle) and $t \bar{t} h$ (right). 
}
\label{fig:smdistris}
\end{figure*}

In the fully leptonic channel,
the main background is the SM
$p p \rightarrow t\bar{t}Z$ production from proton
collisions which 
\TB{we simulate} at leading order
with leptonic decays of the top quarks,
$t \rightarrow b \ell \nu_\ell$, and
$Z \rightarrow \ell^+ \ell^-$.\footnote{We discuss
the background rate normalisation in \cref{sec:numerical}.}
By extracting the $\chel$ and $\chan$ observables,
we show the two-dimensional differential distribution
in the left plot of \cref{fig:smdistris}
in arbitrary units~(a.u.), \ps{defined as the number of events in 
each bin divided by the total number of events.} 
For comparison we additionally show
in the middle plot
the SM distribution of $p p \rightarrow t \bar{t}$
with leptonic decays that 
shows the opposite behaviour,
as expected from the discussion in \cref{sec:angular},
due to the opposite signs of the diagonal elements of
the spin-correlation matrix~\cite{Ravina:2021kpr}.
It should be noted that the 
differences in these distributions
can be attributed to the emission of a spin-one
particle. Unlike $t\bar{t}Z$, the emission of a
spin-zero boson, such as the Higgs boson, does not
induce the same effect.
For comparison,
the distribution for $t \bar{t} h$
production is also shown in
the right plot of \cref{fig:smdistris}\fa{, which has the same overall shape as the $t\bar t$ distribution. }

\subsubsection{Signal}
\label{sec:signalsim}

\begin{figure*}[t]
\centering
\includegraphics[width=0.98\columnwidth]{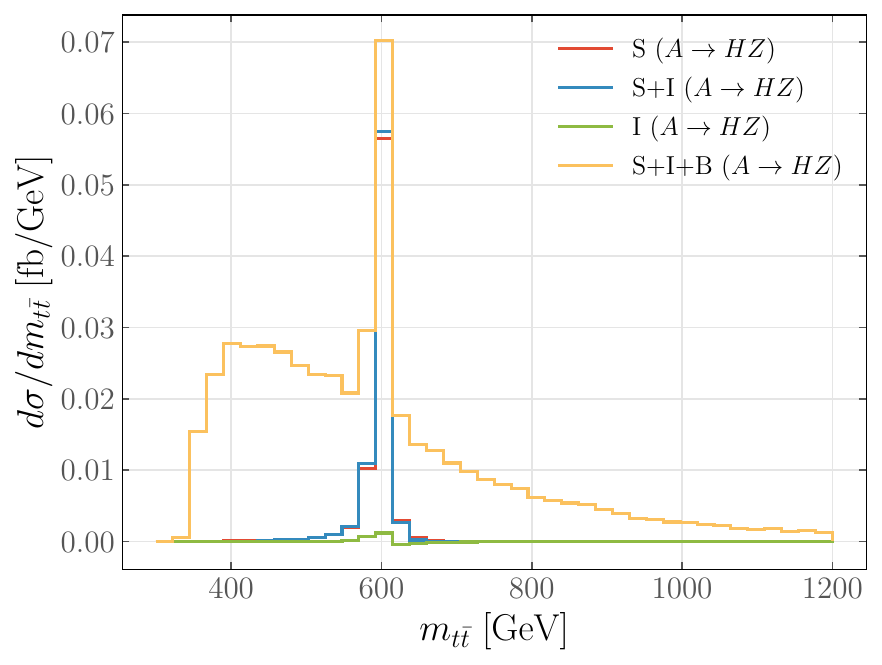}
\includegraphics[width=0.98\columnwidth]{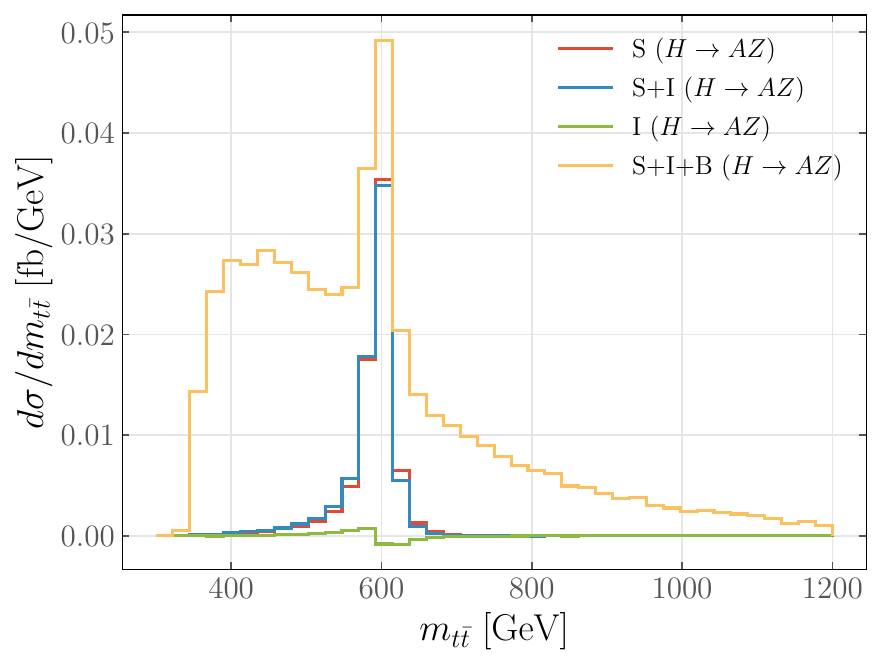}
	\caption{Histograms for the $m_{t \bar t}$ 
        \GW{invariant mass distribution}
    for the $A \rightarrow H Z$ (left) and \GW{the} $H \rightarrow A Z$ process (right). The pure-signal and pure-background contributions are indicated with $S$ and $B$, respectively, while $I$ denotes the signal-background interference at 
        \GW{leading order.}}
\label{fig:interferences}
\end{figure*}

The signal processes under consideration are
\begin{equation}
	\label{eq:proc}
gg \to
\left(
\begin{array}{c}
 A \\
 H
\end{array}
\right) \to
\left(
\begin{array}{c}
 ZH \\
 ZA 
\end{array}
\right) \to
Z \, t \bar t \to
\ell^+ \ell^-  \,
b \bar{b} \ell^+ \ell^- \nu_\ell \bar{\nu}_\ell
\end{equation}
\GW{where we focus in particular on}
the distinction between the $A \rightarrow H Z$ and $H \rightarrow A Z$
signals. The cross sections are corrected by calculating K-factors \GW{as the ratio of 
the QCD next-to-next-to-leading order
$gg \rightarrow A / H$ 
production cross sections 
obtained with 
the {\textsc{HiggsTools}} framework~\cite{Bahl:2022igd},
which incorporates predictions obtained
with \textsc{SusHi}~\cite{Harlander:2012pb,
Harlander:2016hcx},
divided by}
the leading-order $gg \rightarrow A / H$ production cross sections 
\GW{obtained with}
\madgraph.\footnote{The
K-factor for the pseudoscalar is 2.04, while for the scalar it is 2.08 \ps{using a fixed renormalisation scale in \madgraph}.}
As discussed above,
\GW{for the two benchmark scenarios defined in
\cref{sec:2hdm}
the two signal processes $gg \to A \to ZH$ and 
$gg \to H \to ZA$ have 
the same total cross section. This makes}
them indistinguishable in the searches recently
performed by ATLAS and CMS,
where the fully leptonic channel
and the angular information of \GW{the}
final-state leptons
have not been 
\GW{exploited.}

We include interference effects between the signal
and the SM $gg \rightarrow t \bar{t} Z$ 
background and calculate the $\chel$ and $\chan$
observables 
\GW{after subtracting the SM background.}
In particular, differential distributions for the
$g g \rightarrow t \bar{t} Z$ channel
\TB{(including the decays of the top quarks)}
are obtained by calculating the full 
\GW{result for the}
distribution 
\GW{consisting of}
the BSM, SM and interference contributions $d\sigma_\text{full}$ and subtracting
the pure-SM contribution, i.e.\ $d \sigma_{\text{BSM}} =
d \sigma_{\text{full}} - d\sigma_{\text{SM}}$.\footnote{In
practice we achieve this by modifying the matrix element
in \madgraph\ in order to improve numerical stability.}
The interference effects are in general small as shown 
in \cref{fig:interferences}, where
the invariant $m_{t\bar t}$ 
distribution
for the $A \to ZH$ channel 
\GW{is displayed}
on the left and \GW{for}
the $H \to ZA$ channel on the right. 

\TB{We do not include additional
interference contributions between
$g g \rightarrow A/H \rightarrow (H/A) Z$ and box
diagrams resulting in the \GW{same final} 
states,
$g g \rightarrow (H/A) Z$ (an example diagram
is shown in \cref{fig:boxdiag} \GW{below}). 
We have 
checked their importance at the $t \bar{t} Z$
parton level
(not including the decays of the top quarks)
with a loop-ready model
file produced with
{\textsc{NLOCT}}~\cite{Degrande:2014vpa}
and found that the impact of the
box-diagrams is negligible,
see \cref{app:box}.}

\begin{figure*}
\centering
\includegraphics[width=0.98\columnwidth]{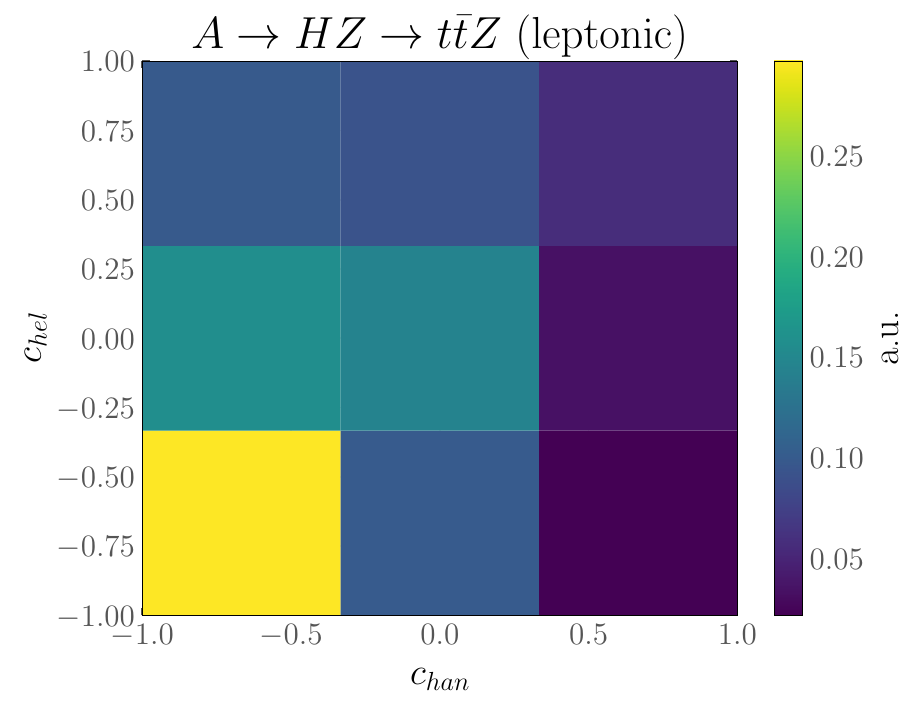}
\includegraphics[width=0.98\columnwidth]{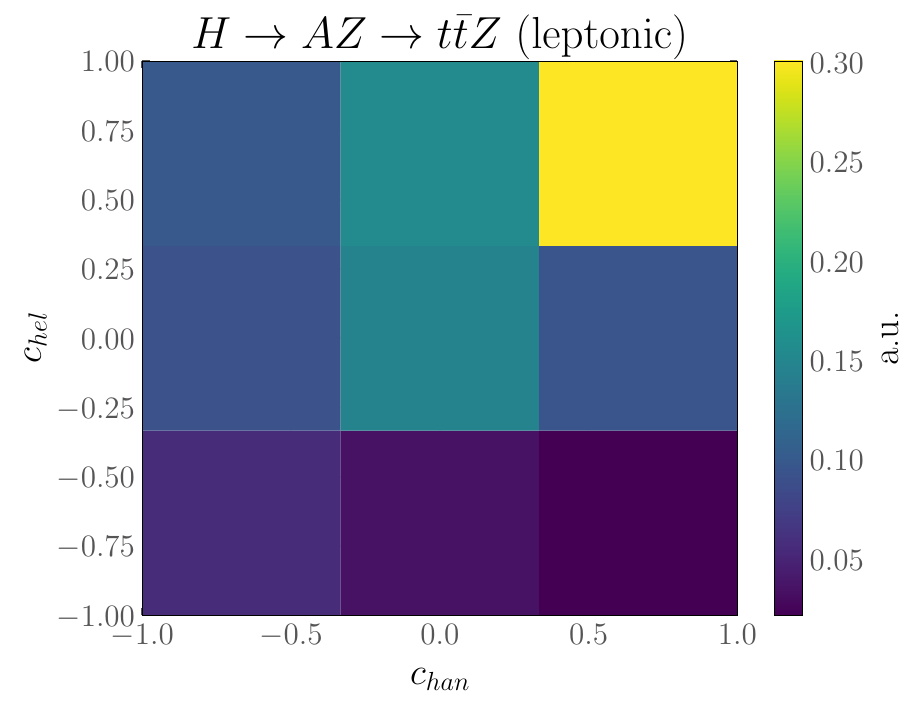}
\caption{Two-dimensional distributions for $\chel$ and $\chan$ for the $A \rightarrow HZ$ and $H \rightarrow AZ$ channels, indicating the potential power to discriminate the two signatures 
\GW{if}
the two observables are utilised in the $t \bar{t} Z$ final state. }
\label{fig:2hdmdistris}
\end{figure*}

The differential cross section distributions in terms of $\chan$ and $\chel$ (in a.u.) for 
\GW{the two} signals $A \rightarrow H Z$ and $H \rightarrow A Z$ are shown in \cref{fig:2hdmdistris} (the range of the colour coding is different than in \cref{fig:smdistris}). 
\GW{The displayed results show that the $\chel$ and $\chan$ variables can potentially provide sensitivity for distinguishing the two signals. While the $A \rightarrow H Z$ signal}
peaks for negative $\chel$ and $\chan$, 
\GW{the $H \rightarrow A Z$ signal} 
\fa{peaks} for positive \fa{$\chel$ and $\chan$}.
\fa{Furthermore, unlike the SM distributions, the  $A \rightarrow H Z$ and $H \rightarrow A Z$ signal distributions are not 
\GW{only}
concentrated in the diagonal bins.} 
\GW{Thus, exploiting the variables}
$\chel$ and $\chan$ 
\GW{and their interplay appears to be a promising approach towards a possible distinction between the two signals}
in a realistic analysis.

\section{Numerical results}
\label{sec:numerical}
We proceed to examine the statistical significance of the two 
considered signals, $A\to HZ$ and $H\to AZ$,
through an analysis utilising the $\chel$ and $\chan$ observables as well as the invariant mass of the di-top state, $m_{t \bar t}$. We design our \GW{phenomenological} 
analysis following the approach of ATLAS for the differential cross section measurements of $t \bar{t} Z$ production~\cite{ATLAS:2023eld}. For simplicity we work with parton-level events, but set requirements for the signal region similar to the ATLAS experiment and subsequently apply Gaussian smearing (see the discussion in \citere{Anuar:2024qsz}).
Selected leptons are required to have a transverse momentum $p_T(\ell) > 10$~GeV and pseudorapidity $\abs{\eta(\ell)} < 2.5$. At least two pairs of opposite-sign same-flavour leptons must be identified, 
\GW{where}
the leading lepton 
\GW{needs to have a}
$p_T$ 
\GW{value}
exceeding $27$~GeV. One \GW{lepton} pair must have an invariant mass close to the mass of the $Z$ boson, $\abs{m_Z -m_{\ell\ell}} < 20$~GeV.  We require at least two jets with $p_T(j) > 25$~GeV and $\abs{\eta(j)} < 2.5$, as $b$-quarks can only be tagged in the central part of the detector. The parton-level top quarks and their daughter leptons are identified from MC-truth information saved in the LHE~\cite{Alwall:2006yp} files. 
\GW{In an actual experimental analysis,}
\ps{the four-momenta of the two top-quarks 
\GW{are reconstructed}
using the four-momenta of the daughter leptons. The proper reconstruction of the $t \bar{t}$-system with fully leptonic decays is 
\GW{non-trivial}
due to the undetected neutrinos, and relies on algebraic kinematic reconstruction methods by imposing $p_T$ conservation and the masses of the $W$-bosons and top-quarks as constraints~\cite{D0:1997pjc,CMS:2012hkm,CMS:2015rld}. The reconstruction efficiency using this technique 
\GW{has been shown to be about} $94\%$ in the $t\bar{t}$ channel~\cite{CMS:2015rld}.}
\psalt{For the $t \bar t Z$ final state, the presence
of the leptonically decaying $Z$ boson does not
hamper the reconstruction
efficiency of the top-quark pair, as its invariant mass
can be identified from a reconstructed
$\ell^+ \ell^-$ pair. Moreover, we verified that
a smearing on the momenta of the top and anti-top
quarks to account for finite detector resolution
effects on the reconstruction has no significant
impact on the $\chel$ and $\chan$ distributions,
see \cref{app:smear}.
We therefore regard it as
sufficient to consider a smearing on the $m_{t \bar t}$
distribution in our analysis.}

ATLAS~\cite{ATLAS:2023eld} expects about $101$ events from the SM $t \bar{t} Z$ channel, while the total \GW{number of} expected events including additional backgrounds rises to $139$ 
\psalt{(the sum of background events in the
SR-4$\ell$-SF and SR-4$\ell$-DF signal regions)}
\GW{for}
an integrated luminosity of $140$/fb. We choose to normalise our background sample such that we obtain $139$ events at the same 
\GW{integrated}
luminosity
(\TB{
we use \madgraph\ to
obtain} the shape of the background distributions,
\TB{assuming that it follow\fa{s}} 
the main $t\bar{t}Z$ 
background). 

Apart from the K-factors applied to both of the signals, as discussed in \cref{sec:signalsim}, we additionally apply efficiency factors 
that 
\GW{equally reduce} the cross section rates 
\GW{for} $A \rightarrow H Z$ and $H \rightarrow A Z$. We apply an efficiency factor of \GW{$(0.7)^2$} for $b$-tagging and a factor of $0.9$ for correctly identifying the reconstructed top quarks and their daughter leptons~\cite{CMS:2012hkm,CMS:2015rld}.\footnote{\ps{The ATLAS analysis only requires one tagged $b$-jet~\cite{ATLAS:2023eld}.}}

Throughout this section we will evaluate \GW{the} statistical significance of the $A\to HZ$ and $H\to AZ$ signals w.r.t.\ the SM in each bin $i$ with 
\begin{equation}
\label{eq:significance}
    Z_i = \sqrt{2 \left[ (S_i+B_i) \log\left(1 + \frac{S_i}{B_i}\right) - S_i\right]}\;,
\end{equation}
where $S_i$ and $B_i$ denote the signal and SM background events in bin $i$, respectively~\cite{Cowan:2010js}. Combined significances are subsequently obtained by summing in quadrature $Z = \sqrt{\sum_i Z_i^2}$ \ps{(without including bin-by-bin correlations)}.
\GW{It should be noted}
\TB{
that we do not include systematic
uncertainties \GW{beyond efficiency factors for $b$-tagging and top-quark reconstruction}, 
\GW{and therefore}
a real experimental analysis is
expected to yield smaller significances than
the ones obtained in our numerical analysis,
see also \GW{the} discussion below.
Nevertheless, our evaluation of the 
\GW{significances will be useful in order}
to quantify the improvement of the
experimental sensitivity 
\GW{as a consequence of incorporating the information from}
$t \bar t$ spin correlations.}



\subsection{$m_{t \bar t}$ distributions}
\label{sec:distributions}

We first investigate the $m_{t \bar t}$ 
distributions from the $A \rightarrow H Z$ channel for a parameter point with $m_A = 650$~GeV and $m_H = 450$~GeV with $\tan\beta = 1$. 
\GW{This is motivated by the fact that ATLAS has observed a 
$2.85\,\sigma$ excess for these mass values, compatible with this 
$\tan\beta$ value, in their search based on the full Run~2 
data set in} 
the semi-leptonic top-quark decay channel~\cite{ATLAS:2023zkt}.
Using bins of $50$~GeV and assuming a $20\%$ Gaussian smearing to approximate detector effects, we evaluate the significance for each bin of the $m_{t \bar t}$ distribution. After summing in quadrature we obtain a combined significance of $3.8\,\fa{\sigma}$ at $140$/fb. While this significance is not directly comparable to the one obtained by ATLAS, as 
\GW{our phenomenological}
analysis \GW{has a different setup and uses} 
a different channel, 
\GW{we regard the fact that the significance that we obtain for this example point is not far away from the ATLAS result as reassuring regarding the validity of our projections for the HL-LHC. The feature that our obtained significance is somewhat higher than the one found by ATLAS is expected since, as discussed above, we neglect systematic effects.}

\begin{figure}
\centering
\includegraphics[width=0.45\textwidth]{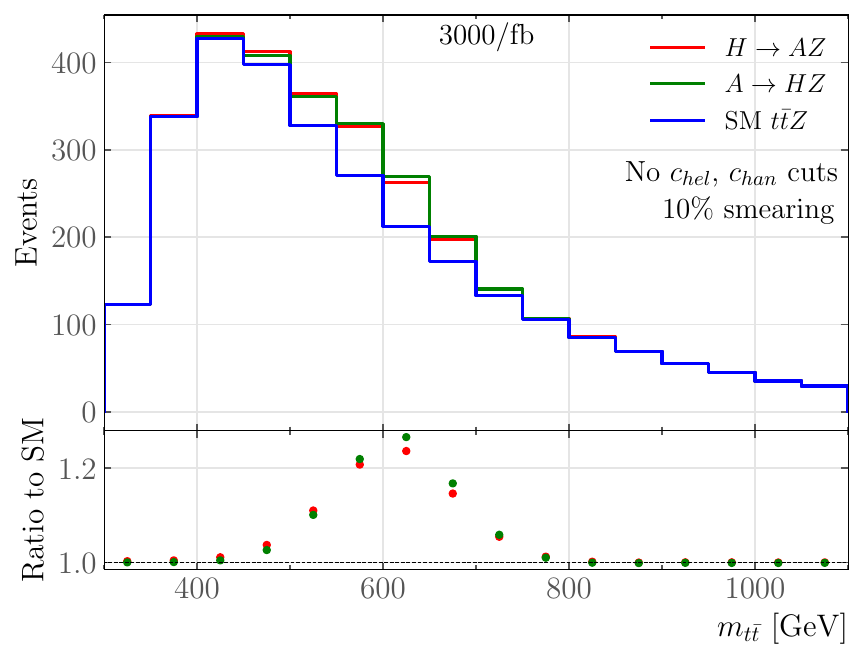}
\caption{\GW{Expected events (upper plot) and ratio to the SM prediction (lower plot) for the di-top invariant mass distribution $m_{t \bar t}$ for an assumed integrated luminosity of $3$/ab
at the HL-LHC. 
The SM contribution to $t\bar{t}Z$ is shown in blue, while the 
$H \rightarrow A Z$ ($A \rightarrow H Z$) signals are shown in
red (green).} 
}
\label{fig:mtt}
\end{figure}

Subsequently, we proceed to study the $m_{t \bar t}$ distributions for $H \rightarrow A Z$ and $A \rightarrow H Z$  for the 
\GW{two benchmark scenarios defined in \cref{sec:2hdm}
(as shown in \cref{fig:entryplot}}
\TB{these parameter points are compatible with the
current experimental limits from LHC searches in
the $Z t \bar t$ final state)}.
\GW{The expected events for the $m_{t \bar t}$ invariant mass distribution are shown
in \cref{fig:mtt} for an integrated luminosity at the HL-LHC of $3$/ab.}
We have assumed an improved smearing of
$10\%$ for the HL-LHC stage.
\fa{As expected}, the $m_{t \bar t}$ distribution does not yield any important differences between the scalar and the pseudoscalar production modes, implying that a resonance search utilising only $m_{t \bar t}$ would be unable to identify the \fa{ CP properties of the resonance state}. Evaluating the significances in each bin according to \cref{eq:significance} and combining 
\GW{the significances for the different bins}
yields a significance of \ps{$5.9$ ($5.5$) for $A \rightarrow H Z$ ($H \rightarrow A Z$) with the assumed $10\%$ smearing effect.}

\subsection{Discrimination \GW{between $A \to ZH$ and $H~\to~ZA$}}
\label{sec:discri}

\begin{figure*}[t!]
\centering
\includegraphics[width=0.9\textwidth]{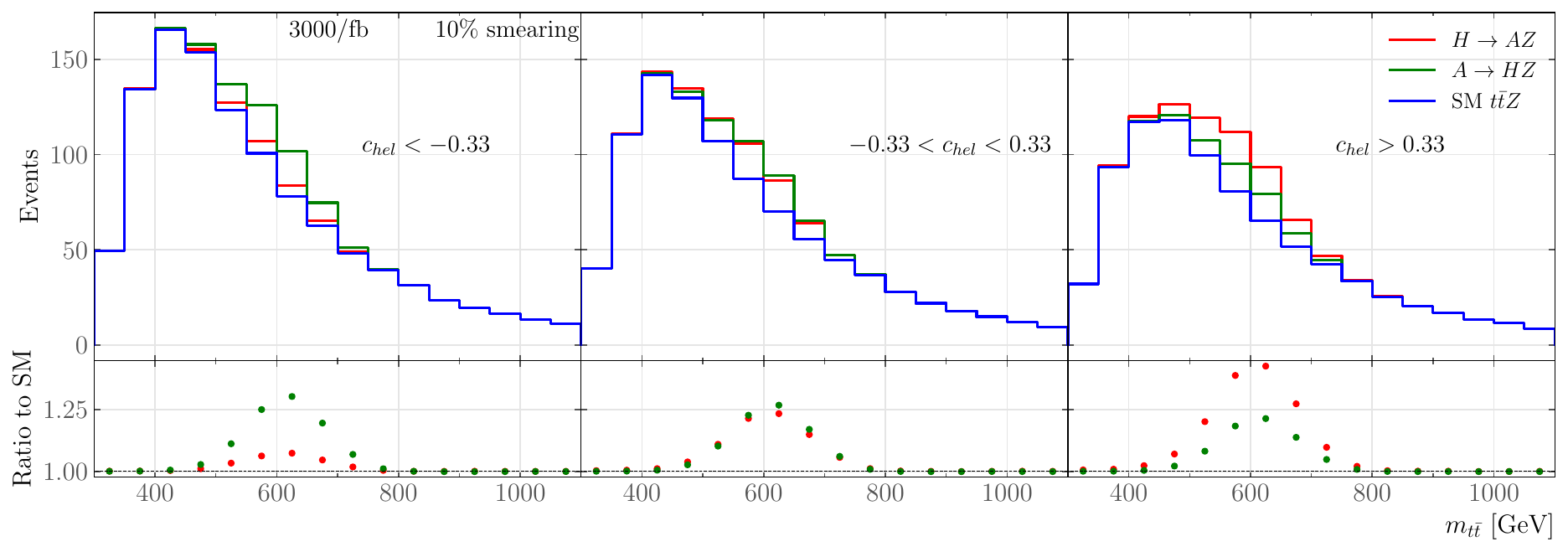}
\includegraphics[width=0.9\textwidth]{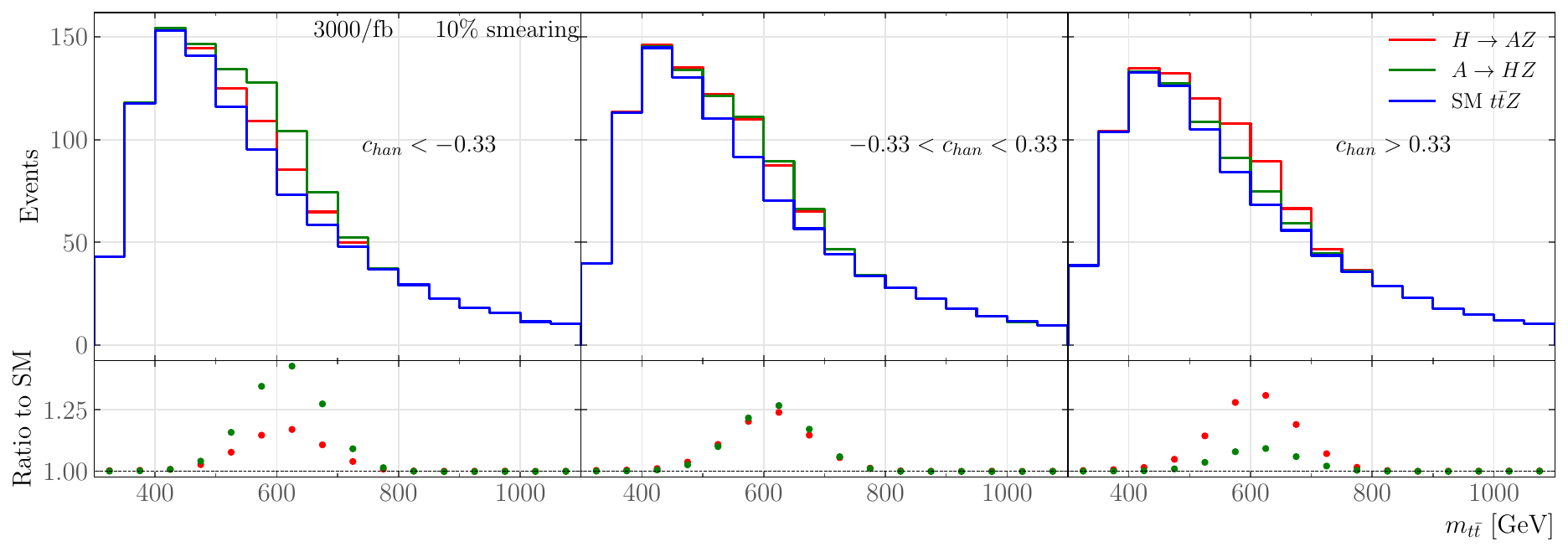}
\caption{\GW{Expected events (upper plots) and ratio to the SM prediction (lower plots) for the di-top invariant mass distribution $m_{t \bar t}$ for different $\chel$ (top) and $\chan$ (bottom) regions 
for an assumed integrated luminosity of $3$/ab
at the HL-LHC. 
The SM contribution to $t\bar{t}Z$ is shown in blue, while the 
$H \rightarrow A Z$ ($A \rightarrow H Z$) signals are shown in
red (green).}
}
\label{fig:chel_or_chan_hllhc}
\end{figure*}

We \GW{now incorporate information from $t \bar t$ spin correlations. As a first step we consider the case where either the angular observable $\chel$ or $\chan$ is utilised. As a possible binning in $\chel$,}
the generated events can be separated into the three different regions $\chel < -0.33$, $-0.33 < \chel < 0.33$ and $\chel > 0.33$ 
\GW{such that separate results are obtained for}
the $m_{t \bar t}$ distributions in each region. 
\GW{For the case where $\chan$ is utilised as single angular observable the same kind of binning can be chosen.
Results for the two case\ps{s} are shown in \cref{fig:chel_or_chan_hllhc}, where the expected} 
$m_{t \bar t}$ distributions \GW{of the events and the ratio of the distributions including the $H \to AZ$ or $A \to HZ$ signals to the SM prediction are displayed} for the different regions 
\GW{in 
$\chel$ or $\chan$}
\fa{for an integrated luminosity of}
$3$/ab. 
\GW{The choice of using
three regions for $\chel$ or $\chan$ rather than a finer binning was made in order to ensure that the resulting $m_{t\bar{t}}$ 
bins} are not depleted of background events. 
The $m_{t \bar t}$ binning of $50$~GeV has been kept as \GW{for the case of \cref{sec:distributions}
where no angular variable is utilised}.
In line with the expectations from \cref{fig:2hdmdistris}, 
\GW{for the case where $\chel$ is utilised as angular variable} one observes a higher ratio of $A \rightarrow H Z$ events with respect to the SM for the $\chel < -0.33$ \GW{region}. 
\fa{In contrast,} $H \rightarrow A Z$ yields a higher ratio \fa{w.r.t.\ the SM} for $\chel > 0.33$. The highest value of the ratio across all regions is \GW{obtained} for the $H \rightarrow A Z$ signal in this case. A similar pattern can also be observed for the $\chan$ regions, albeit in this case the ratio to the SM across all the regions reaches higher values for $A \rightarrow H Z$. This indicates that $\chel$ and $\chan$ have discriminating power allowing the separation 
\GW{between the $A \rightarrow H Z$ and $H \rightarrow A Z$ signals, 
in contrast to the case where}
one uses only the cross section rates or only the $m_{t \bar t}$ distribution.

\begin{figure*}[t]
\centering
    \includegraphics[width=0.45\textwidth]{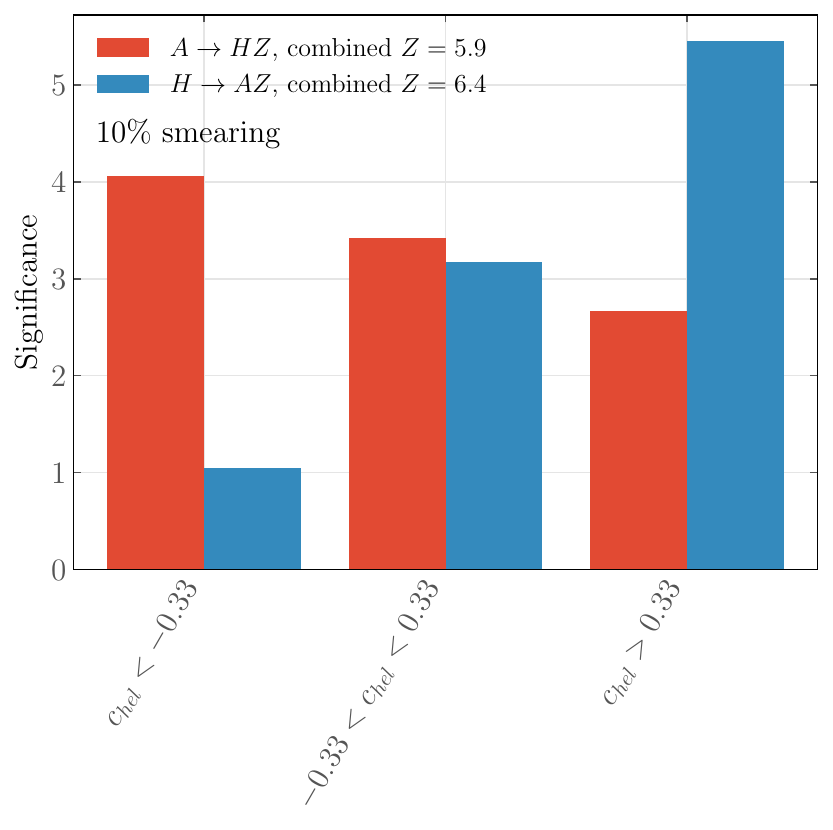}
    \includegraphics[width=0.45\textwidth]{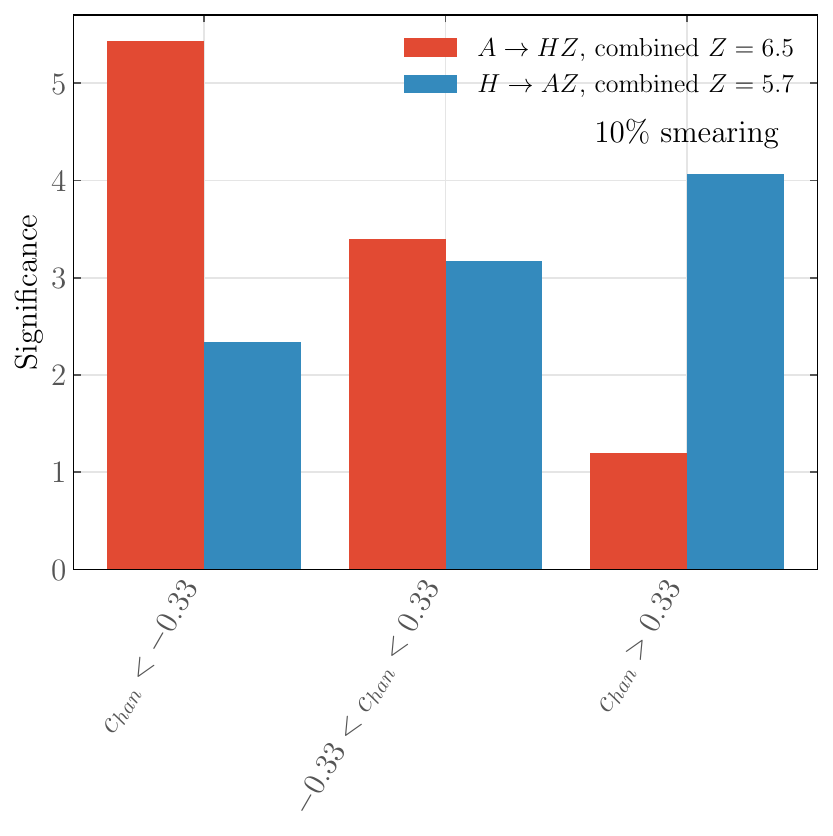}
\caption{On the left, the significances 
\GW{that are obtained for the two signals}
are shown for the analysis splitting the events only according to \GW{the angular variable} $\chel$. On the right we show the significances for the analysis that 
\GW{utilises only the angular variable}
$\chan$. 
}
\label{fig:sensitivities_hllhc}
\end{figure*}

\GW{In order to further quantify the sensitivity to the two signals, we evaluate}
the \fa{statistical} significance in each $m_{t \bar t}$ \fa{ distribution
\GW{according to}
\cref{eq:significance}} and then 
\GW{obtain the combined significance for each bin of the angular observable. The results are displayed in
\cref{fig:sensitivities_hllhc}}.
\GW{We find that both cases, i.e.\ utilising only the angular variable $\chel$ or only the angular variable $\chan$,}
yield high significances for both \GW{signals} 
$H \rightarrow A Z$ and $A \rightarrow H Z$. 
\GW{For the two  cases the combined significances follow a similar pattern across the three bins of the angular variable.}
\TB{Furthermore,} 
\GW{the two types of signals can be distinguished and thus,
assuming that a potential signal consists of a CP eigenstate,
the CP nature of this state can be determined. If in this case} the highest significance 
\GW{is obtained from utilising a binning in $\chel$ and arises} 
\TB{mostly from events with} $\chel > 0.33$,
then evidently the signal is due to the production of a 
\GW{CP-even state,} $g g \rightarrow H$. The opposite is true for a signal that originates from a 
\GW{CP-odd state,} $g g \rightarrow A$,
\GW{for which the
highest significance would occur in the region} 
$\chan < -0.33$. 

\GW{We furthermore note}
that utilising the angular observables has the potential to yield a higher significance compared to simply using the $m_{t \bar t}$ distribution. 
\TB{We find that the significance of the two signals
does not surpass $6\,\sigma$ 
\GW{for the case where}
the spin-correlation
observables are not 
\GW{taken into account}.
In contrast, binning in $\chel$ \GW{($\chan$)}
gives rise to a significance of more than $6\,\sigma$
for the $H \to AZ$ \GW{($A \to HZ$)} signal, 
as shown in
\cref{fig:sensitivities_hllhc}.
We checked that this remains true even if one increases
the number of bins in the $m_{t \bar t}$ distribution (we tried this for two cases
by reducing the bin-size to 25~GeV and to 16.7~GeV).}
\GW{We note that this improvement of the significance of a detected signal will only occur in the data set where both top quarks decay leptonically, while signal regions based on semileptonically or hadronically decaying top-quark pairs will not be affected.}

\begin{figure*}
\centering
	\includegraphics[width=0.52\textwidth]{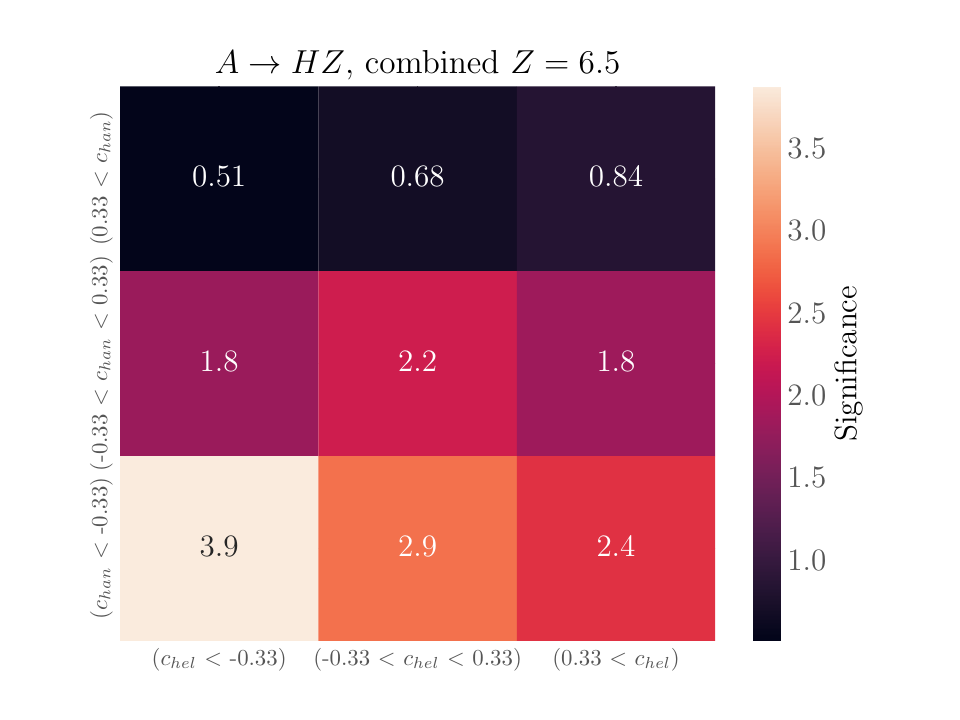}\hspace{-1cm}
    \includegraphics[width=0.52\textwidth]{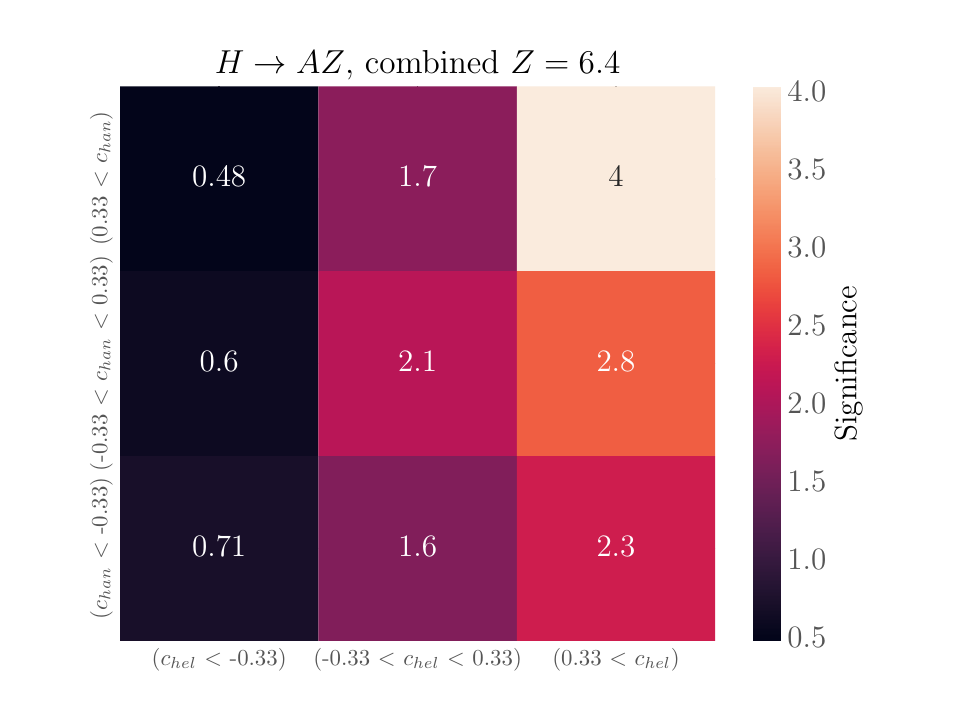}
	\caption{Significances for the approach \GW{where a simultaneous binning in} both $\chel$ and $\chan$ \GW{is} utilised to define different regions. On the left we show the results for $A \rightarrow H Z$ and on the right for $H \rightarrow A Z$ 
        \GW{for} $3$/ab \ps{with $10\%$ smearing}. 
 }
\label{fig:sensitivities_2dregs}
\end{figure*}

\GW{We now turn to the case where the two angular observables are exploited simultaneously. Thus, instead of}
\TB{only binning in either $\chel$
or $\chan$}, 
\GW{we now}
use nine different
regions \GW{arising from a simultaneous}
binning in both $\chel$ and $\chan$ 
and construct $m_{t \bar t}$ distributions within each region. 
\GW{For this purpose we readjust}
the number of bins in the $m_{t \bar t}$ distribution to $80$~GeV in order to avoid depleting any 
bin of background events and compute the combined significances in each region for the 
\GW{$H \to AZ$ and $A \to HZ$ signals.
These results} are shown in \cref{fig:sensitivities_2dregs}. 
\GW{We find}
similar conclusions to the
\GW{case where only one of the angular observables is employed. 
The simultaneous binning in both observables yields a high sensitivity for distinguishing between the 
$A \to HZ$ and $H \to AZ$ signals, where the former is expected to have the highest significance in the $(\chel < -0.33, \, \chan < -0.33)$ bin while the latter is expected to occur with the highest significance in the $(0.33 < \chel, \, 0.33 < \chan)$ bin.
The overall significances for each of the two signals are found to be similar as for the case where only one of the angular observables is employed if}
the significances 
in the different regions are combined. Ultimately, the 
appropriate \GW{choice for the binning in the angular observables and the $m_{t \bar t}$ distribution will} 
depend on the number of events that 
\GW{are obtained in the actual analysis at the HL-LHC}.

\section{Summary and conclusions}
\label{sec:conclusions}

In this work, we have focused on (HL-)LHC
searches for a new spin-0 boson, produced through gluon fusion,
that decays into a $Z$ boson and a lighter new spin-0 boson 
of opposite 
\GW{CP character, where
the latter decays} into a pair of top quarks.
\GW{Such a signal}
\TB{has been identified as a \GW{``smoking-gun''}
signature for a first-order EW phase transition
in \GW{extended Higgs sectors like} 
the Two Higgs doublet model~(2HDM). 
In the absence of CP violation, 
\GW{the two} spin-0 resonances 
\GW{involved in this process will be a CP-even state, $H$, and a CP-odd state, $A$, which can occur either as parent or daughter particle in the decay. 
Thus, the process can give rise to}
two possible signals,
$gg \to A \to ZH \to Z t \bar t$ and
$gg \to H \to ZA \to Z t \bar t$.}
The current experimental searches
performed by ATLAS and CMS lack sensitivity
to the CP properties of the spin-0 bosons.
Therefore, 
\GW{these searches}
cannot distinguish between 
\GW{the two kinds of}
possible signals
if the total production rates
for the $Zt\bar t$ final state are predicted to be 
\GW{the same within the involved theoretical and experimental uncertainties}.

We 
\GW{have proposed}
here a method to distinguish between
the $A \to ZH$ and $H \to ZA$ signals that
exploits the spin correlations
of the $t \bar t$ system.
To achieve this, we 
utilise the
observables $\chan$ and $\chel$
(defined in \cref{eq:chanhel}), which are
reconstructed from the angular distributions
of the two leptons produced in the fully
leptonic decays of the \GW{two} top quarks. 
These observables were previously 
applied in LHC searches \GW{for a single new resonance} in the $t\bar t$ channel.
\GW{We have demonstrated that the application of these angular observables can successfully be extended to signatures involving two BSM particles.
In particular we have shown that exploiting the}
$t \bar t$ spin correlations is
valuable in the $Zt\bar t$ channel 
\TB{both for determining the CP properties
of new spin-0 resonances and to increase the overall
experimental sensitivity.}

We \GW{have used} the CP-conserving
2HDM as a
\TB{minimal UV-complete } 
BSM framework
that predicts two neutral \TB{BSM} 
Higgs bosons $H$ and $A$, which are CP-even and CP-odd, respectively.
Therefore, our analysis then focuses on discriminating between
the signals
\TB{$H \to ZA$ and $A \to ZH$,
which is phenomenologically of great interest
since the presence of the latter is more favourable
for a realisation of a strong first-order EW
phase transition in the 2HDM. Such a \GW{phase} transition
is required for an explanation of
the observed matter-antimatter asymmetry
of the universe via EW baryogenesis, and it gives
rise to a primordial gravitational-wave background
that might be detectable with future space-based
gravitational-wave experiments, such as LISA.}

For our numerical analysis, we \TB{have defined} 
two benchmark 
\GW{scenarios} \TB{featuring} 
a \TB{resonant $Z t \bar t$} production cross section of
0.1~pb and masses of 800~GeV and 600~GeV for the two Higgs bosons. 
While these rates are currently
\TB{below the experimental sensitivity of} 
the \GW{searches at the LHC in the $Zt\bar t$ channel,
they are expected to be} within the reach of the HL-LHC.
\GW{In a first step,}
we have evaluated the statistical significance of the $A\to HZ\to t\bar tZ$ and $H\to AZ\to t\bar tZ$ signals from the cross section distributions w.r.t.\ the top-quark pair invariant mass, $m_{t\bar t}$.
We have 
\GW{then} explored the impact of considering the spin correlations from the di-top system by considering 
\GW{either a binning for one of the angular observables
$\chel$ or $\chan$ or a simultaneous binning for both of the angular observables}.

Our results show that the 
signal $A \to HZ$ \GW{yields contributions to the angular observables that 
predominantly occur in the regions where $\chel$ and $\chan$ are negative}, while for $H \to AZ$ 
\GW{the largest contributions occur in the regions
where both $\chel$ and $\chan$ are positive}.
This behaviour allows \GW{one} to infer 
\GW{information about}
the CP nature of the 
\GW{produced states, which is not possible with the}
current search strategy \TB{applied by ATLAS and CMS.
In this way, a distinction between the two
signals 
\GW{$A \to HZ$ and $H \to AZ$}
is possible with high statistical significance,
\GW{which has important implications for assessing}
the possible realisation of EW baryogenesis 
\GW{and the prospects for}
a future detection of gravitational waves with 
\GW{observatories such as}
LISA.}

\GW{It should be noted that}
\TB{the realisation of EW baryogenesis also
requires new sources of CP violation.
Therefore, in the future
it would be worthwhile to investigate how precisely
the CP properties of the 
\GW{produced new states}
could be
determined if they are not CP eigenstates but
a mixture of CP-even and CP-odd components.}
\GW{Such an analysis is beyond the scope of the present paper.}

We have also shown that 
\GW{applying a binning of events with respect to}
either one 
\GW{of the} angular variables 
\GW{(with an appropriate choice of $\chel$ or $\chan$ for the two signals) or both of them}
yields a higher statistical significance for the signals compared to the SM background 
for both $A \to HZ$ and $H \to AZ$. 
For our chosen benchmark \GW{scenarios}, we find an enhanced statistical significance of 6.4--6.5 
\GW{for the case where the information from the angular variables $\chel$ and $\chan$ is included,}
compared to \ps{5.5--5.9}
\GW{for the case where this information is not taken into account}.
\TB{Hence, the separation of events in different bins
of $\chan$ and $\chel$ 
is also promising 
\GW{for increasing}
the overall
experimental sensitivity of the experimental searches,
irrespective of whether the signature originates
from the $A \to ZH$ or the $H \to ZA$ decay.}
\GW{It should be kept in mind, however, that our proposed analysis strategy is only directly applicable to the data set where both top quarks decay leptonically.}

\faalt{While the significances
obtained in an actual experimental analysis may turn out to be somewhat smaller (due to
systematic uncertainties not considered in our analysis) or larger (due to other systematic
improvements in future CMS and/or ATLAS analyses) compared to our estimates, our results
for the relative differences, which are the basis for demonstrating the discriminating power of our
method, should provide reliable information about the potential of employing the spin
correlation variables.
In addition, since we have accounted for several sources of experimental uncertainties in our analysis, such as smearing of the $m_{t\bar{t}}$ distributions, our results should be unlikely to be significantly altered by detector effects.}

In summary, we find that 
\GW{exploiting the information from}
top-quark spin correlations, through the observables $\chel$ and $\chan$, can provide crucial 
\GW{sensitivity to}
the CP properties of BSM 
\GW{states that are} detectable in current and future $Zt\bar t$ searches at the (HL-)LHC, 
\GW{while the}
current searches 
\GW{carried out by ATLAS and CMS are insensitive to the CP nature of the produced states}. 
\TB{We encourage the experimental
collaborations to 
\GW{make use of this important source of information in their}
future Higgs-boson searches in the
$Z t \bar t$ final state (and other $t\bar{t}$+$X$ final states) at the LHC} 
\GW{by improving their search strategies along the lines that have been proposed in this paper}.

\section*{Acknowledgements}
\psalt{We thank Matthias Schr\"oder
and Laurids Jeppe for helpful
discussions.}
The project that gave rise to these
results received the support of a
fellowship from the ``la Caixa''
Foundation (ID 100010434). The
fellowship code is  LCF/BQ/PI24/12040018.
T.B.~acknowledges the support of the Spanish Agencia
Estatal de Investigaci\'on through the grant
``IFT Centro de Excelencia Severo Ochoa CEX2020-001007-S''.
F.A., P.S.\ and G.W.\ acknowledge support by the Deutsche
Forschungsgemeinschaft (DFG, German Research
Foundation) under Germany's Excellence
Strategy -- EXC 2121 ``Quantum Universe'' --
390833306.
The work of F.A., P.S.~and~G.W.~has also been partially funded
by the Deutsche Forschungsgemeinschaft 
(DFG, German Research Foundation) -- 491245950.

\appendix

\section{Impact of box-diagram contribution}
\label{app:box}

\begin{figure}
\centering
\includegraphics[width=0.6\columnwidth]{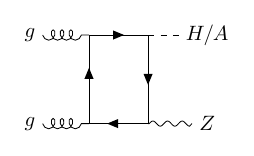}
\caption{Box diagram resulting in the 
production of a scalar $H$ or pseudoscalar $A$
in association with a $Z$ boson.
}
\label{fig:boxdiag}
\end{figure}

\begin{figure*}[t]
\centering
\includegraphics[width=0.98\columnwidth]{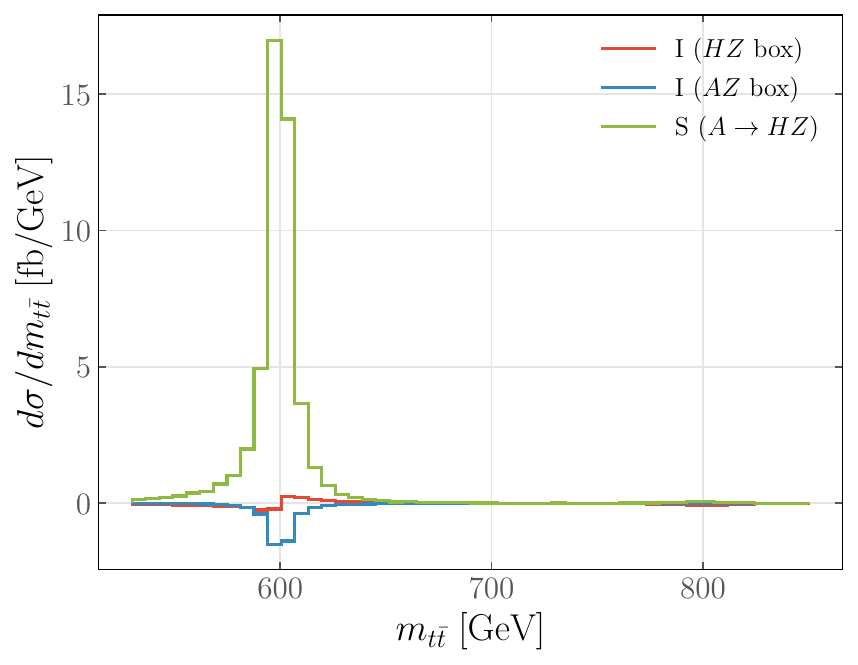}
\includegraphics[width=0.98\columnwidth]{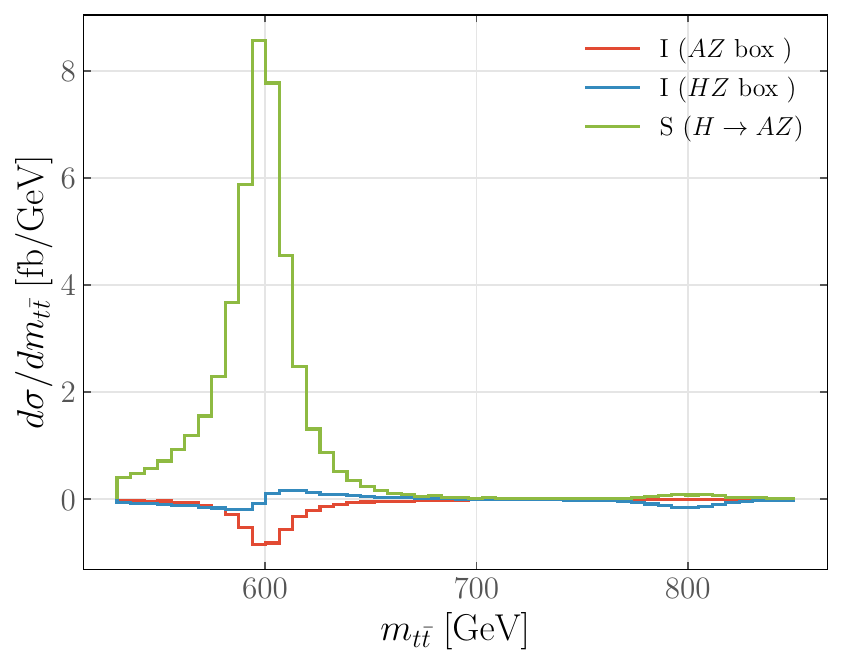} \\
\includegraphics[width=0.98\columnwidth]{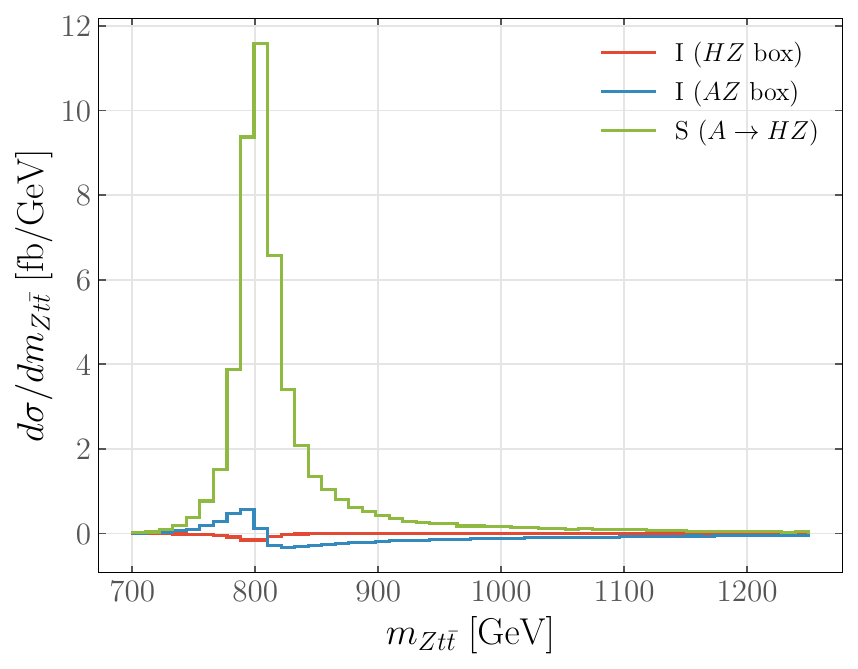}
\includegraphics[width=0.98\columnwidth]{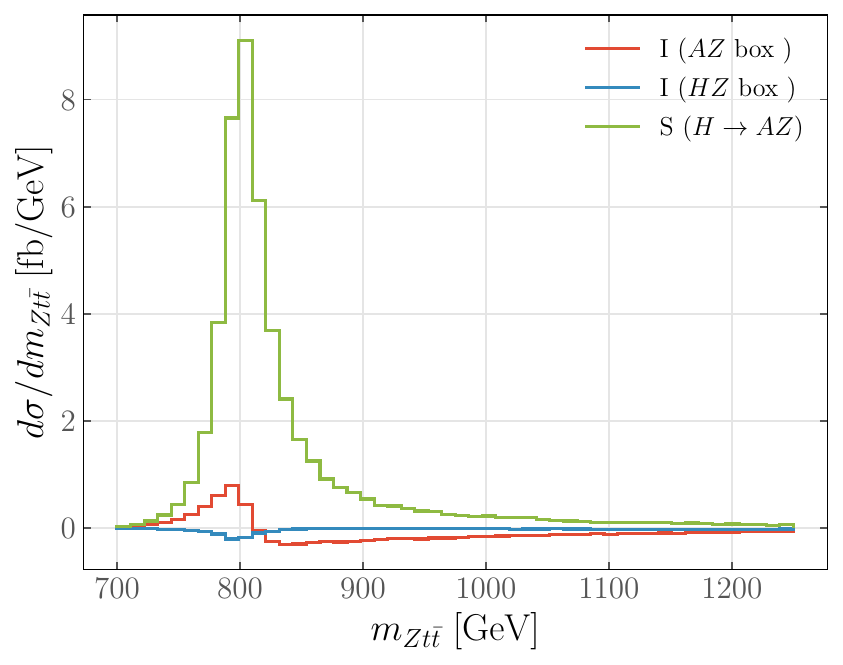} 
\caption{Distributions for the invariant mass of the \GW{top-quark pair}, $m_{t \bar t}$, (top) and \GW{for} the total invariant mass of the process, $m_{Zt \bar t}$, (bottom). On the left (right) the pure signal contribution, $\mathrm{S}$, to the $A \rightarrow H Z$ ($H \rightarrow A Z$) channel \GW{is shown, where $H$ ($A$) decays to top quarks. The} interference between \GW{the} signal and box contributions is denoted \GW{by}~$\mathrm{I}$. }
\label{fig:interferences_box}
\end{figure*}

The $ZH$ and $ZA$ final states considered here
\GW{can also be} produced 
via \GW{contributions involving a} box-diagram.
The corresponding \GW{type of} Feynman diagram
\GW{is} depicted in \cref{fig:boxdiag}.
In order to 
\GW{assess}
the relevance of
these contributions,
we simulated events 
with \madgraph~including the box-diagrams
at the $t \bar{t} Z$
parton level
(not including the decays of the top quarks)
with a loop-ready model
file produced with
{\textsc{NLOCT}}~\cite{Degrande:2014vpa}.
In \cref{fig:interferences_box} we show the cross
section distributions for the $A \to ZH$ signal
(left) and the $H \to ZA$ signal (right)
as a function of the invariant
mass of the top-quark pair (top row) and the invariant
mass of the $t \bar{t} Z$ state (bottom row)
for different contributions: the total signal
\GW{(green)}, the interference between the resonant production
and the box-diagram
leading to $HZ$ production \GW{(red)}, and the
interference between the resonant production and the
box-diagram leading to $AZ$ production \GW{(blue)}.
Overall, 
we find that
the interference effects (blue and red lines)
are small compared
to the signal 
(green line).
Our findings are
in agreement with \citeres{Goncalves:2022wbp,
ATLAS:2023zkt,
CMS-PAS-B2G-23-006}.
We therefore do not take into account the 
production of a spin-0 resonance in association with
a $Z$-boson via the box diagrams 
\GW{in}
our Monte-Carlo simulations.

\section{\psalt{Impact of smearing top-quark momenta on $\chel$ and $\chan$}}
\label{app:smear}

\begin{figure*}[h!]
\centering
    \includegraphics[width=0.98\columnwidth]{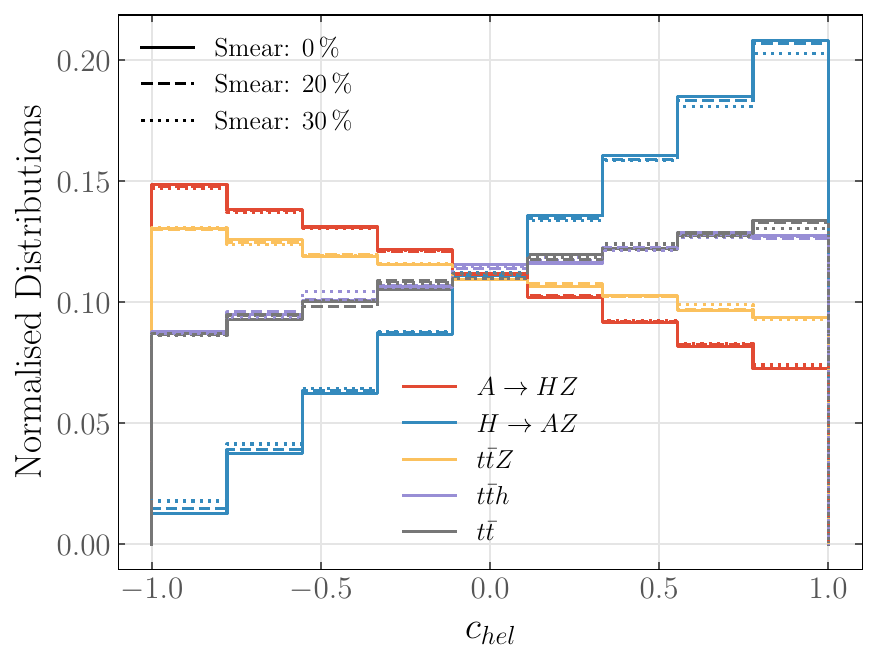}
    \includegraphics[width=0.98\columnwidth]{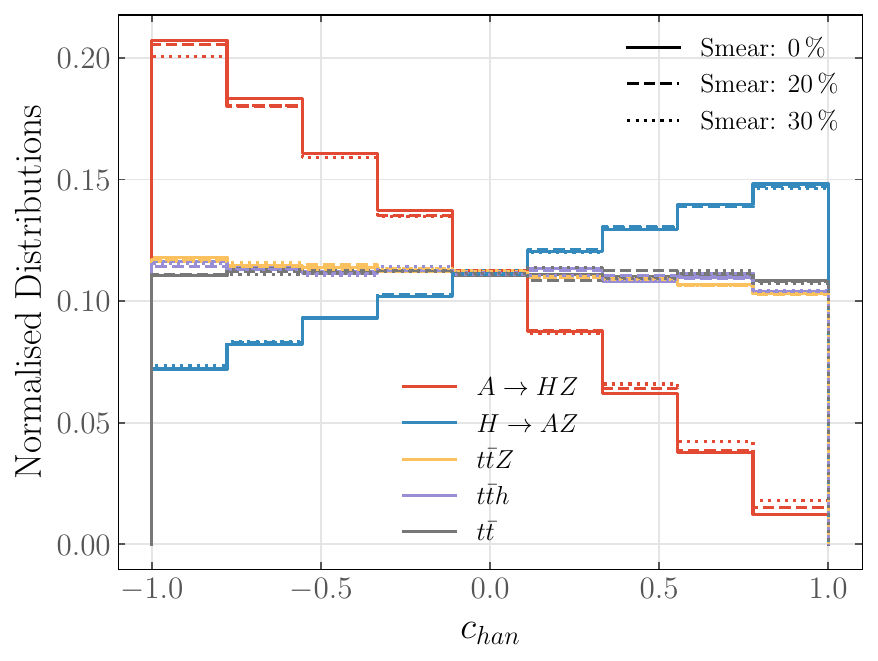} 
\caption{\psalt{Distributions for $\chel$ and $\chan$ with $0\%$, $20\%$ and $30\%$ smearing on the momenta of the MC-truth top and anti-top quarks. The $t \bar{t} Z$ background is shown, as well as the $A \rightarrow H Z$ and $H \rightarrow A Z$ channels with the lighter scalar decaying to $t\bar{t}$. For comparison, we also show distributions for $t\bar{t}$ and $t\bar{t}h$.}}
\label{fig:smeared_chel_chan}
\end{figure*}

\psalt{In experimental analyses, the need to reconstruct the top-quark momenta in final states involving neutrinos through algebraic reconstruction methods can lead to uncertainties in the components of the reconstructed momenta with respect to the actual momenta of the top quarks. This is not taken into account in our analysis since we use the MC-truth top-quark momenta and only apply a reconstruction efficiency to accommodate the case where top quarks could not be reconstructed. 
Here we investigate how sensitive the $\chel$ and $\chan$ observables are to shifts of the components of the top-quark momenta caused by the finite detector resolution.
To simulate detector effects,
we perform a conservative $20\%$ and $30\%$
Gaussian smearing on the $p_x,p_y,p_z$ components of the four-momenta for both the top and anti-top quarks and subsequently re-compute the $\chel$ and $\chan$ variables for the events in our signal and background samples. We note that we do not smear the mass of the individual top quarks since it is usually kept fixed when experiments perform the algebraic reconstruction of the momenta~\cite{CMS:2015rld}. The $\chel$ and $\chan$ distributions for $A \rightarrow HZ \rightarrow t \bar{t}Z$, $H \rightarrow AZ \rightarrow t \bar{t}Z$ and $t \bar{t} Z$ are shown in \cref{fig:smeared_chel_chan} (we additionally show the distributions for $t\bar{t}$ and $t \bar{t} h$ for comparison). Overall, the impact from such shifts of the top-quark four-momenta is found to be small and we therefore do not include these effects in our analysis.}

\bibliographystyle{JHEP}
\bibliography{refs}

\end{document}